\documentclass[11pt,aps,preprint,nofootinbib]{revtex4}
\usepackage[active]{srcltx}
\usepackage[utf8]{inputenc}
\usepackage{latexsym}
\usepackage{amsmath}
\usepackage{graphicx}
\usepackage{yfonts}
\usepackage{pgfplots}

\begin{document}

\title{Spontaneous Breaking of Lorentz Symmetry with an antisymmetric tensor}

\author{C. A. Hernaski}
\email{carlos.hernaski@gmail.com}
\affiliation{Departamento de F\'isica, Universidade Estadual de Londrina, \\
Caixa Postal 10011, 86057-970, Londrina, PR, Brasil}

\begin{abstract}

Spontaneous violation of Lorentz symmetry by the vacuum condensation of an antisymmetric $2$-tensor is considered. The coset construction for nonlinear realization of spacetime symmetries is employed to build the most general low-energy effective action for the Goldstone modes interacting with photons. We analyze the model within the context of the Standard-Model Extension and noncommutative QED. Experimental bounds for some parameters of the model are discussed, and we readdress the subtle issues of stability and causality in Lorentz non-invariant scenarios. Besides the two photon polarizations, just one Goldstone mode must be dynamical to set a sensible low-energy effective model, and the enhancement of the stability by accounting interaction terms points to a protection against observational Lorentz violation. 

\end{abstract}
\maketitle
\section{Introduction}

Constraints due to Lorentz invariance are strong enough to allow any significant deviations of this symmetry. Indeed, 
there are severe restrictions on the possible sizes of Lorentz-violating parameters in a number of models \cite{Russell}.
Even so, the study of theories without Lorentz symmetry is believed to be relevant for understanding physics at extremely high energies. Some approaches to tackle the problem of the quantization of gravity like string theory, loop quantum gravity, and warped brane worlds seems to have enough room to allow possible Lorentz symmetry violations at some energy scale and it has been the focus of numerous investigations in the literature \cite{lqg,st,wb}. At the same time, other approaches directly affect the very notion of a continuous spacetime like noncommutative geometry, which introduces a special tensor simulating the spacetime non-smoothness and leading to Lorentz symmetry breaking \cite{ncg}. In any case, experimental searches for Lorentz-violating signals at low energies have been attained Planck scale sensitivity \cite{Russell}, setting up the possibility of probing energy scales where quantum gravity plays an important role.

From the theoretical perspective, the connivance with Lorentz violation in standard quantum field theory brings about some concerns regarding established useful properties granted by exact Lorentz symmetry. In this respect we are mainly interested with the issues of stability and micro-causality in Lorentz-violating models. In particle physics language stability demands the absence of ghosts and tachyons in the particle spectrum, whereas microcausality requires that observables commute for spacelike separations. The former aims to ensure vacuum stability, while the later is a way to implement the classical notion of causal chain of events in spacetimes that only allow finite velocity signals. Both are expected properties of sensible models and are worthy to be investigated in general since in the relativistic context these notions are closely intertwined with Lorentz invariance.

In the work of Ref. \cite{alan-ralf}, the authors analyze stability and causality by considering the Dirac Lagrangian for massive spin $1/2$ particles in the presence of constant background fields in flat spacetime, which explicitly breaks Lorentz invariance. The upshot is that for sufficiently small Lorentz-violating backgrounds, stability and causality can be simultaneously assured up to some characteristic energy scale. Above that scale some difficulties are brought to the fore concerning either stability or causality. Even if Lorentz transformations acting on the particles or localized fields are not an invariance of the model, there is still an observer Lorentz invariance that can be implemented to simplify the analysis. However, the lack of stability or causality is an observer independent notion according to discussion in Ref. \cite{alan-ralf}. It is also argued that even considering the model with explicit symmetry breaking as a sector of a fully Lorentz-invariant Lagrangian that triggers the violation spontaneously, hardly would change the situation. In Ref. \cite{carroll}, a similar consistency analysis is employed in a vector model that breaks Lorentz symmetry spontaneously. By demanding stability in every frame, restrictions are obtained on the form of the dispersion relations of the free particles in agreement with the conclusions of Ref. \cite{alan-ralf}.

In the present work we revisit the discussion of stability and causality in connection with spontaneous Lorentz symmetry violation. To this end, we consider a specific model where an antisymmetric $2$-tensor develops a non-vanishing expectation value in vacuum through some unknown underlying mechanism. We concentrate the analysis in the low-energy regime of the model, so the massive modes that possibly can be generated at the phase transition and the very details of the underlying mechanism itself can be safely ignored. Our motivation to consider such a model is to make contact with noncommutative field theories. 

In this context, it is argued that spacetime itself loses its meaning when we consider short-distance behavior. 
A basic argument is that the spacetime exhibits a kind of discrete structure, which can be simulated by means of a noncommutativity 
between the spacetime coordinates \cite{Doplicher},
\begin{equation}
[x^{\mu},x^{\nu}]=i\theta^{\mu\nu},
\label{1.1}
\end{equation}
where $\theta^{\mu\nu}$ is a real and antisymmetric constant matrix.

A Lagrangian for noncommutative fields can be rewritten in terms of commutative ones by means of the Moyal product, defined by
\begin{equation}
f(x)\star g(x)\equiv \exp\left(\frac{1}{2}i\theta^{\mu\nu}\partial_{x^\mu}\partial_{x^\nu}\right)f(x)g(y)\Big|_{x=y}.
\label{1.2}
\end{equation}
In this way, it is possible to treat any realistic Lagrangian coming from noncommutative geometry as a special case of the Standard-Model extension (SME) \cite{sme,alan-ncg}, with $\theta^{\mu\nu}$ being in this case the only background tensor field that originates Lorentz-violating operators. In Ref. \cite{alan-ncg}, this identification is applied to noncommutative QED, which renders an upper bound on some of the $\theta^{\mu\nu}$-parameters of the order of $(10\ \ TeV)^{-2}$.

In the forthcoming sections we will investigate the possibility of allowing fluctuations of the $\theta$-parameters around their constant value. Namely, we will consider the constant $\theta$-matrix as resulting from the vacuum condensation of an antisymmetric $2$-tensor through some underlying mechanism in a fully Lorentz-invariant model. So, besides the usual explicit Lorentz-violating effects originated from the usual noncommutative scenarios, we will consider contributions of the Nambu-Goldstone (NG) bosons associated to the fluctuations of the background field. Therefore, Lorentz symmetry will be considered as spontaneously broken instead of explicitly violated.

Besides, we will focus on the low-energy degrees of freedom. Instead of considering some specific Lorentz-invariant model exhibiting a detailed mechanism through which the antisymmetric field develops a non-vanishing vacuum expectation value (VEV), we will center the attention only on the broken symmetry pattern and the degrees of freedom that are independent on the specific model we could start with. We understand that this effective field theory approach is more powerful when we do not know the details of the underlying interactions that give rise to the low-energy dynamics and offers a broader framework to probe sensible models. In this sense, our discussion complements a similar earlier work that also considers the spontaneous Lorentz symmetry breaking by an antisymmetric $2$-tensor \cite{alan-ncg}.

A very general and systematic analysis of models exhibiting spontaneous symmetry breaking is obtained through the coset construction, from which we will avail ourselves. In this formalism, the focus is the low-energy regime of the model, and one of its main advantages is for the practical purpose of calculating scattering processes, since the perturbation series is organized in terms of the energy of the processes instead of the coupling constants, as emphasized in \cite{weinberg2}.

To understand what kind of phenomenology the scenario considered in this work could provide, we envisage the coupling of the NG bosons to other Standard-Model fields. The photon sector plays an important role in the low-energy regime where no massive particle can be excited. For this reason we construct the effective Lagrangian of the NG bosons of the antisymmetric field coupled to the photon and investigate the possible departures from the standard QED. Comparing the model with the photon sector of the SME, we derive bounds on the Lagrangian parameters to be consistent with observations. The issues of microcausality and stability are then reassessed, and we argue that both can be maintained in this context. A nonlinear analysis of the stability conditions points to a protection against observational Lorentz violation.

The plan of the paper is as follows: In Sec. \ref{nonlinear}, we review the main points of the coset construction for the nonlinear realization of internal and spacetime symmetry groups. In Sec. \ref{ssb}, the coset construction for the Lorentz group broken to the vacuum invariant subgroup of our interest is considered, and we build the effective action describing the NG modes interacting with photons in a Lorentz-violating environment. In Sec. \ref{stability}, we consider a toy model to address the main subtleties involved in the analysis of causality and stability of Lorentz non-invariant models. Then, in Sec. \ref{causality}, we apply the analysis of Sec. \ref{stability} to our specific model. Finally, in Sec. \ref{sum}, we present our final remarks. In the Appendix we derive the Lorentz-violating dispersion relations of the propagating modes.

\section{Coset construction and nonlinear realizations\label{nonlinear}}

\subsection{Internal symmetries}

We begin this section by reviewing the main points of the coset construction. Let us first consider a physical system described by a general Lagrangian $\mathcal{L}$ invariant under the action of an internal (commuting with the Poincaré group) semisimple Lie group $G$. Also, we suppose the dynamics of the system favors some of the fields to develop a non-vanishing VEV $\langle\phi_{\alpha}\rangle=v_{\alpha}\neq 0$, where $\alpha$ denote that $\phi_{\alpha}(x)$ belongs to some specific representation of $G$. The subset of elements of $G$ that leaves $v_{\alpha}$ invariant forms a subgroup $H$ of $G$. Let us denote by $T_i$ the generators of the subgroup $H$ and by $X_a$ the generators of the group elements of $G$ that act effectively on the vacuum subspace. The generators of the semisimple Lie algebra can be chosen so as to satisfy the Cartan decomposition
\begin{eqnarray}
\left[T_i,T_j\right] & = & if_{ijk}T_k, \label{tt}\\
\left[T_i,X_a\right] & = & if_{iab}X_b, \label{tx}\\
\left[X_a,X_b\right] & = & if_{abc}X_c+if_{abi}T_i. \label{cartan decomposition algebra}
\end{eqnarray}
When $f_{abc}=0$ the coset space $G/H$ is called a symmetric space. Any group element of $G$ can then be decomposed in the following way
\begin{equation}
g=\exp\left(i\omega_aX_a\right)\exp\left(i\theta_iT_i\right). \label{cartan decomposition group}
\end{equation}

Since $\mathcal{L}$ is invariant under $G$, any $\langle\phi_{\alpha}\rangle^g$ obtained from $\langle\phi_{\alpha}\rangle$ through the action of an element $g$ $\in$ $G$, is also a minimum energy configuration. Naively we say that the vacuum is infinitely degenerate, but strictly speaking the distinct vacua cannot belong to the same separable Hilbert space. However, one expects a complete equivalence among all quantum theories obtained by choosing distinct vacua. Any new configuration of the order parameter obtained by the action of some group element can be interpreted as zero momentum massless excitation, since it does not require energy to be excited.

Only elements of the coset space $G/H$ act effectively on the order parameter. According to the decomposition (\ref{cartan decomposition group}), the coset space can be parametrized as $\exp\left(i\omega_aX_a\right)$. So, a finite momentum excitation can be obtained by a local group transformation as
\begin{equation}
\phi_{\alpha}(x)=U\left(\xi(x)\right)_{\alpha\beta}v_{\beta}=\exp\left(i\xi_a(x)X_a\right)_{\alpha\beta}v_{\beta}, \label{coset NG}
\end{equation}
where the local fields $\xi_a(x)$ are the NG bosons. One can clearly see that there is one NG for each broken generator. More generally, the field $\phi_{\alpha}$ contains other degrees of freedom called massive modes besides the NG modes. Denoting these modes by $\chi_{\alpha}(x)$, we can generalize the expression above to 
\begin{equation}
\phi_{\alpha}(x)=\exp\left(i\xi_a(x)X_a\right)_{\alpha\beta}\left(v_{\beta}+\chi_{\alpha}(x)\right). \label{field red}
\end{equation}
Usually the massive fields $\chi_{\alpha}$ satisfy some constraints in order to match the number of degrees of freedom present in $\phi_{\alpha}$. The field reparametrization (\ref{field red}) must be applied to other fields $\psi(x)$ present in the initial Lagrangian by extracting from them the NG degrees of freedom via the redefinition $\psi(x)=e^{i\xi_a(x)X_a}\tilde{\psi}(x)$.

The coset parametrization defines a nonlinear realization of the group $G$. In fact, let us consider the action of an arbitrary element of $G$ on the coset element $\exp\left(i\xi_a(x)X_a\right)$. From (\ref{cartan decomposition group}) we have
\begin{eqnarray}
gU\left(\xi(x)\right) & = &\exp\left(i\xi^{\prime}(x)_aX_a\right)\exp\left(i\mu_i\left(\xi(x),g\right)T_i\right),\nonumber \\
                      & = & U\left(\xi^{\prime}(x)\right)h\left(\xi(x),g\right).\label{coset trans1}
\end{eqnarray}
The transformation of the NG fields are in general nonlinear, but when restricted to elements of the subgroup $H$ one gets a linear representation
\begin{equation}
\xi^{\prime}_a(x)=h_{ab}\xi_b(x).
\end{equation}
In the same way, from a matter field $\psi(x)$ transforming linearly under $H$, we obtain the field $\tilde{\psi}(x)$, which transforms nonlinearly under full group $G$ as
\begin{equation}
\tilde{\psi}^{\prime}_a(x)=\left(e^{i\mu_i\left(\xi(x),g\right)T_i}\right)_{ab}\tilde{\psi}_b(x),
\end{equation}
with the same element of $H$ that appears in the transformation rule for the coset element in (\ref{coset trans1}).

If we start with a Lagrangian with fields in the linear representations of the broken symmetry and then we perform the previous field redefinitions, we obtain a Lagrangian that is invariant under the same group $G$, but only with $H$ acting linearly on the fields. Since the original $\mathcal{L}$ is invariant under global $G$ transformations, the only terms where a NG field can manifest are the ones containing derivatives. Specifically, the NG fields only appear through the combination $U^{-1}(x)dU(x)$, called Maurer-Cartan $1$-form, which belongs to the Lie algebra of $G$ and plays an important role in the coset construction framework. We define the geometrical objects $D_{\mu}\xi_a(x)$ and $\omega_{\mu a}$ by
\begin{equation}
U^{-1}(x)dU(x)=idx^{\mu}\left(X_aD_{\mu}\xi_a(x)+T_i\omega_{\mu i}(x)\right).\label{MC form}
\end{equation}
It is straightforward to show the transformations rules for $D_{\mu}\xi_a(x)$ and $\omega_{\mu a}(x)$. Under a general transformation of $G$, we have
\begin{eqnarray}
D_{\mu}\xi^{\prime}_a(x) & = & \mathcal{D}_{ab}\left(h(\xi(x),g)\right)D_{\mu}\xi_b(x), \label{NG cov der trans}\\
\omega^{\prime}_{\mu i}(x) & = & \mathcal{E}_{ij}\left(h(\xi(x),g)\right)\omega_{\mu j}(x)-\mathcal{H}_{ib}\left(h(\xi(x),g)\right)\partial_{\mu}\xi_b(x), \label{con trans1}
\end{eqnarray}
where
\begin{eqnarray}
hT_ih^{-1} & = & i\mathcal{E}_{ij}(h)T_j,\\
\left(\partial_{\mu}h\right)h^{-1} & = & T_i\mathcal{H}_{ib}(h)\partial_{\mu}\xi_b(x).
\end{eqnarray}

From (\ref{NG cov der trans}), we see that $D_{\mu}\xi_a(x)$ transforms covariantly under the action of a general element $g$ of $G$. For this reason it is called the NG covariant derivative. On the other hand, $\omega_{\mu i}$ transforms as a connection and can be used to construct covariant derivatives of matter fields, as can be readily shown from the transformation rule of the quantity
\begin{equation}
\nabla_{\mu}\tilde{\psi}(x)\equiv \partial_{\mu}\tilde{\psi}(x)+T_i\omega_{\mu i}(x)\tilde{\psi}(x), \label{cov mat}
\end{equation}
which is the same as $\tilde{\psi}(x)$.

The Maurer-Cartan form (\ref{MC form}) provides all the elements that have a simple transformation law under the nonlinear realization of the broken symmetry. In this way, in an effective approach where we are mainly interested in the low-energy regime of the model, we can use these elements to construct invariant Lagrangians from the onset. The main advantage of this procedure is the independence of the underlying mechanism through which the symmetry is spontaneously broken. In fact the only information we need to construct the effective Lagrangian is the broken symmetry pattern $G\longrightarrow H$. Furthermore, as seen above, the NG fields always enter in the Lagrangian with a derivative, and this allows a perturbative expansion of the scattering amplitudes in terms of the characteristic energy of external NG particles. Hence, even when the complete Lagrangian is known and the fields are in the linear representation of the symmetry group, in the low-energy regime, it may be convenient to perform the field redefinitions to recast the Lagrangian with fields in the nonlinear realization of the broken symmetry due to the suitability of the NG energy expansion.

\subsection{Spacetime symmetries}

We will apply the ideas discussed above to the case of spacetime symmetries. These are defined in general as transformations on the fields that do not commute with the Poincaré generators. The coset formalism for spacetime symmetries was mainly developed in \cite{volkov} and can be reviewed in \cite{ogievetsky}. A nice discussion of the coset construction in connection with the SME can be found in \cite{penco}. The main subtlety that must be accounted is the special role played by translations. Even when the translations are not broken, they act nonlinearly on the spacetime coordinates as $x^{\prime\mu}=x^{\mu}+a^{\mu}$. In this way, the spacetime coordinates $x^{\mu}$ can be thought as parameters of the coset space (Poincaré)/(Lorentz). The symmetry group $G$ includes the unbroken translation generators $P_{\alpha}$, the other unbroken generators $T_{i}$ of a subgroup $H$, and the broken generators $X_a$. We emphasize that both, $T_i$ and $X_a$, in general contain internal and spacetime generators. Following \cite{volkov}, the coset $G/H$ should be parametrized as
\begin{equation}
U\left(\xi(x),x\right)=e^{ix^{\mu}P_{\mu}}e^{i\xi_a(x)X_a}. \label{coset spacetime}
\end{equation}
Notice the special place occupied by the unbroken translations. The factor $e^{ix^{\mu}P_{\mu}}$ ensures the correct transformation of the coordinates under spacetime symmetries. Under an arbitrary transformation $g$ of $G$, one gets
\begin{equation}
gU\left(\xi(x),x\right)=U\left(\xi^{\prime}(x^{\prime}),x^{\prime}\right)h\left(\xi(x),g\right), \label{coset trans2}
\end{equation}
with $h\left(\xi(x),g\right)$ being an element of $H$, but depending on the NG fields and the $g$ element.

From (\ref{coset spacetime}) we can calculate the Maurer-Cartan $1$-form, which is the basic structure that provides the fundamental building blocks for the construction of invariant effective Lagrangians. We have
\begin{equation}
U^{-1}dU=idx^{\mu}\left(e_{\mu}^{\ \alpha}P_{\alpha}+X_aD_{\mu}\xi_a+T_i\omega_{\mu i}\right). \label{MC spacetime}
\end{equation}
In addition to the analogous structures $D_{\mu}\xi_a$ and $\omega_{\mu i}$ that already appeared in the calculation of the Maurer-Cartan form for the internal case, we also have the extra component $e_{\mu}^{\ \alpha}$ defined via $e_{\mu}^{\ \alpha}P_{\alpha} \equiv g^{-1}P_{\mu}g$, which, as we will see in more detail in the following section, plays the role of a vielbein, mapping objects in the nonlinear realization to corresponding objects in the linear one. From (\ref{coset trans2}) we can get the transformation rules for the components of Maurer-Cartan form. Defining $e_P\equiv dx^{\mu}e_{\mu}^{\ \alpha}P_{\alpha}$, $D\xi\equiv dx^{\mu}X_aD_{\mu}\xi_a$, and $\omega\equiv dx^{\mu}T_i\omega_{\mu i}$, we find
\begin{eqnarray}
e^{\prime}_P & = & he_Ph^{-1},\\
D\xi^{\prime} & = & hD\xi h^{-1},\label{cov der trans}\\
\omega^{\prime} & = & h\omega h^{-1}+hdh^{-1}. \label{con trans2}
\end{eqnarray}

The first two objects transform covariantly under arbitrary $G$ transformations and can be used directly in the construction of invariant Lagrangian, whereas the connection $\omega$ is necessary to construct covariant derivatives for the matter fields. For  a matter field $\psi(x)$ belonging to a matrix representation $\mathcal{D}$ of $H$, we define the transformation under an arbitrary $G$ by
\begin{equation}
\psi^{\prime}(x^{\prime})=\mathcal{D}\left(h\left(\xi(x),g)\right)\right)\psi(x).
\end{equation}
We can then define the covariant derivative of $\psi$ by
\begin{equation}
\nabla\psi(x)=d\psi(x)+T_i\omega_i\psi(x),
\end{equation}
which transforms in the same way as $\psi(x)$.

To get a explicit nonlinear description of the effective Lagrangian, it is desirable to express the tensor quantities in a noncoordinate basis, which transforms nonlinearly under $G$, contrary to the basis inherited by the coordinates $x^{\mu}$. For this purpose it is convenient to define the inverse of the vielbein $e^{\mu}_{\ \alpha}$ by the relation
\begin{equation}
e^{\mu}_{\ \alpha}e_{\mu}^{\ \beta}=\delta^{\alpha}_{\beta}, \label{viel1}
\end{equation}
and then perform the change of basis in the tangent space:
\begin{equation}
e_{\alpha}=e^{\mu}_{\ \alpha}\partial_\mu. \label{viel2}
\end{equation}
We can then extract the fully covariant structures $D_{\alpha}\xi_b$ and $\omega_{\alpha i}$ from the components of the Maurer-Cartan form, $dx^\mu D_{\mu}\xi_b$ and $dx^\mu\omega_{\mu i}$, in the following way
\begin{eqnarray}
D_{\alpha}\xi_b & = & e^{\mu}_{\ \alpha}D_\mu\xi_b,\\
\omega_{\alpha i} & = & e^{\mu}_{\ \alpha}\omega_{\mu i}.
\end{eqnarray}

To construct the invariant action, we take the covariant quantities with noncoordinate basis indices and contract them with the metric in the same noncoordinate basis:
\begin{equation}
g_{\alpha\beta}\equiv e^{\mu}_{\ \alpha}e^{\nu}_{\ \beta}\eta_{\mu\nu}.
\end{equation}
The possibly non-trivial vielbein determinant must also be accounted for the invariance of the integration measure. This is given by
\begin{equation}
\text{det}(e_\mu^{\ a})d^4x.\label{measure}
\end{equation}

From our discussion, the action
\begin{equation}
S=\int\text{det}(e_\mu^{\ a})d^4x\mathcal{L}\left(D_{\alpha}\xi_b,\nabla_{\alpha}\psi,g_{\alpha\beta}\right)
\end{equation}
will be invariant under nonlinear transformations of the group $G$.

To end this review section, we need to mention the so-called {\it inverse Higgs effect} \cite{inverse-HE}. This is another important subtlety that nonlinear realization of spacetime symmetries can manifest. As discussed in \cite{inverse-HE}, in the case of broken spacetime symmetries, the counting of independent NG modes is not as direct as in the internal case, where one mode is associated to each broken symmetry generator. Following \cite{manohar}, it may happen that linearly independent broken generators are related through commutation with translation generators. In this case, the NG modes associated to these generators, which are local symmetry transformations of the order parameter, will not be independent. That is, if $X_i$ and $X_j$, with $i\neq j$, are two broken generators, and if $[P_\mu,X_i]\sim X_j+\dots$, the NG modes associated to $X_i$ and $X_j$ will not be independent from each other. One can then eliminate the $X_j$-related NG mode by setting the $X_j$ component of the Maurer-Cartan form to zero. This constraint is invariant under $G$, since the Maurer-Cartan form is itself invariant, and can be consistently implemented. As will become clear in the next section, this effect is absent in our particular discussion and the naively NG counting will in fact apply.

\section{Spontaneous violation of Lorentz symmetry \label{ssb}}

We will contemplate the possibility of a Poincaré invariant model induce spontaneous Lorentz violation by the vacuum condensation of an antisymmetric $2$-tensor $\Theta_{\mu\nu}(x)$. Directly related to this scenario are the noncommutative models, in which the Poincaré degeneracy is lifted by the presence of the noncommutative parameters $\theta_{\mu\nu}$, defined in (\ref{1.1}), contracted with the dynamical fields. In some sense, we will consider an extension of these models by allowing fluctuations of the noncommutative parameters. The strict connection of our discussion with noncommutative field theories is obtained with identification $\langle\Theta_{\mu\nu}(x)\rangle=\theta_{\mu\nu}$. The focus will be in the low-energy regime of the model; {\it i.e}, far below the scale $\Lambda_{NC}$, where spontaneous symmetry breaking occurs and is supposed to be near the Planck Mass, $M_P\sim 10^{19}\ GeV$. In fact, the low-energy noncommutative QED Lagrangian starts to present Lorentz-violating effects with dimension six operators. So, we have a suppression of the order of $(\Lambda_{NC})^2$ compared to the Maxwell term. In our approach, we will consider Lorentz deviations already with marginal operators. This still can be related to noncommutative QED if we consider that this theory has a cutoff $\Lambda$, and the Lorentz-violating marginal operators are produced through radiative corrections of dimension six operators \cite{anisimov}. Then, the parameters of dimension four Lorentz-violating operators are at the order of $\left(\Lambda/\Lambda_{NC}\right)^2$ and can be highly suppressed if we suppose $\Lambda\ll \Lambda_{NC}$.

In the low-energy limit, it is reasonable to integrate out the massive modes generated in the breaking mechanism and consider the effective theory for the NG modes. Following the reasoning of the previous section, we then consider the most general effective Lagrangian compatible with the breaking pattern $G\rightarrow H$, with the initial group $G$ taken as the Poincaré group $PO(1,3)$ and $H$ the Lorentz subgroup that leaves $\theta_{\mu\nu}$ invariant. Since $\theta_{\mu\nu}$ is a constant matrix, the translation part of the Poincaré group remains as an exact symmetry of the model.

Let us define a dimensionless matrix $\bar{\theta}_{\mu\nu}$ through $\theta_{\mu\nu}=1/(\Lambda_{NC})^2\bar{\theta}_{\mu\nu}$. Assuming $\text{det}\left(\bar{\theta}_{\mu\nu}\right)\neq 0$, it is possible to choose a special coordinate system to put $\bar{\theta}_{\mu\nu}$ in the form
\begin{equation}
\bar{\theta}_{\mu\nu}=\left(\begin{array}{cccc}
0 & a & 0 & 0\\
-a & 0 & 0 & 0\\
0 & 0 & 0 & b\\
0 & 0 & -b & 0
\end{array}\right), \label{theta}
\end{equation}
with arbitrary nonzero $a$ and $b$. In this coordinate system the unit vector tangent to the rest observer worldline is given by
\begin{equation}
n_{\mu}\equiv \left(1,0,0,0\right).\label{unit}
\end{equation}
Therefore, we can define the preferred spacelike vector
\begin{eqnarray}
e_{\nu} & \equiv & \frac{1}{n_\rho n_\sigma\bar{\theta}^{\rho}_{\ \lambda}\bar{\theta}^{\lambda\sigma}}n^\mu\bar{\theta}_{\mu\nu},\\
& = & \left(0,1,0,0\right). \label{preferred vec}
\end{eqnarray}
It is then clear from the form of the matrix (\ref{theta}) that any combination of a boost along $e_{\mu}$ with a rotation around this same vector leaves $\bar{\theta}_{\mu\nu}$ invariant. The two kinds of transformations are clearly independent and commute with each other. Therefore, the invariant subgroup $H$ is of the form $SO(1,1)\otimes SO(2)$.

Let us take the Poincaré algebra in $3+1$ spacetime dimensions\footnote{Our metric convention is $\eta^{\mu\nu}=(1,-1,-1,-1)$.}:
\begin{eqnarray}
\left[J^{\mu\nu},J^{\rho\sigma}\right] & = & -i\left(\eta^{\nu\rho}J^{\mu\sigma}+\eta^{\mu\sigma}J^{\nu\rho}-\eta^{\mu\rho}J^{\nu\sigma}-\eta^{\nu\sigma}J^{\mu\rho}\right), \label{jj alg}\\
\left[P^{\mu},J^{\rho\sigma}\right] & = & -i\left(\eta^{\mu\rho}P^{\sigma}-\eta^{\mu\sigma}P^{\rho}\right), \label{pj alg}\\
\left[P^{\mu},P^{\nu}\right] & = & 0 \label{pp alg}.
\label{2.1}
\end{eqnarray}

It is convenient to decompose the Poincaré algebra in terms of irreducible representations of $H=SO(1,1)\otimes SO(2)$. We will use Latin capital letters, $A,\ B,\ C,\ldots$, to denote the representations of the $SO(1,1)$ part of $H$ and lower case second half of the Latin alphabet, $i,\ j,\ k,\ldots$, to denote the representations of the $SO(2)$ part. Without loss of generality, we will take the indices $A,\ B,\ C,\ldots$ to assume values $0$ and $1$ and the indices $i,\ j,\ k,\ldots$ to assume values $2$ and $3$. With these conventions, we can call as $T^{01}$ and $T^{23}$ the unbroken boost and rotation generators of the groups $SO(1,1)$ and $SO(2)$, respectively, and by $X^{01},\ X^{02},\ X^{12},\ X^{23}$ the broken ones. As it is clear, $X^{01}$ and $X^{02}$ represent boosts along the two spacelike directions perpendicular to the $e_{\mu}$ vector (\ref{preferred vec}), and $X^{12}$ and $X^{23}$ represent rotations around these same two directions. This allows us to put the algebra (\ref{jj alg})-(\ref{pp alg}) into the form
\begin{eqnarray}
\left[X^{Ai},X^{Cj}\right] & = & i\left(\lambda^{AC}T^{ij}+\gamma^{ij}T^{AC}\right), \label{xx alg}\\
\left[X^{Ai},T^{CD}\right] & = & i\left(\lambda^{AD}X^{Ci}+\lambda^{AC}X^{iD}\right), \label{xtA alg}\\
\left[X^{Ai},T^{jk}\right] & = & -i\left(\gamma^{ij}X^{Ak}+\gamma^{ik}X^{jA}\right), \label{xti alg}\\
\left[T,T\right] & = & 0,\ \left[X,P\right] \sim P,\ \left[T,P\right] \sim P,\ \left[P,P\right] = 0, \label{rest alg}
\end{eqnarray}
with the $X$'s and $T$'s tensors taken as antisymmetric under their indices exchanges, and $\lambda^{AB}$ and $\gamma^{ij}$ being defined respectively by
\begin{equation}
\lambda^{AB}=\left(\begin{array}{cc}
1 & 0\\
0 & -1
\end{array}\right),\label{lambda}
\end{equation}
and
\begin{equation}
\gamma^{ij}=\left(\begin{array}{cc}
-1 & 0\\
0 & -1
\end{array}\right).\label{gamma}
\end{equation}

By comparing (\ref{xx alg})-(\ref{rest alg}) with (\ref{tt})-(\ref{cartan decomposition algebra}), one can notice that the algebra (\ref{xx alg})-(\ref{pp alg}) is in the Cartan-decomposition form. Also, as we will soon verify, the matrices $\lambda^{AB}$ and $\gamma^{ij}$ play the role of metrics for constructing invariants under nonlinear realizations of $SO(1,3)/H$.

Special attention is drawn to the form of the commutator of a broken generator with the translation generators $[X,P]$. It involves only translation generators, which are unbroken in the present discussion. So, as we anticipated at the end of the last section, there is no need to worry about the {\it inverse Higgs effect}, and the number of NG modes is equal to $4$, the same number as the broken $X$-generators.

To each one of the $X^{Ai}$ generators we associate the corresponding NG field $B_{Ai}(x)$. Like $X^{Ai}$, we take $B_{Ai}$ as antisymmetric. So, as claimed before, there are only $4$ independent NG fields. These fields parametrize the coset manifold $SO(1,3)/\left(SO(1,1)\otimes SO(2)\right)$. Together with the coordinates $x^{\mu}$ associated to the unbroken translations, the full coset $PO(1,3)/\left(SO(1,1)\otimes SO(2)\right)$ is parametrized as
\begin{eqnarray}
U\left(x;B_{Ai}\right)&=&\exp\left(ix^{\mu}P_{\mu}\right)\exp\left(iB_{Ai}\left(x\right)X^{Ai}\right), \nonumber \\
&=&e^{ix^{\mu}P_{\mu}}\Omega\left(B_{Ai}\left(x\right)\right). \label{cos}
\end{eqnarray}

Let $\omega_{\mu\nu}$ and $a^\mu$ be the parameters of arbitrary Lorentz and translation transformations, respectively. Then, under the left action of a general element of $PO(1,3)$, $g(\omega,a)=e^{ia^\mu P_\mu} e^{\frac{i}{2}\omega_{\mu\nu}J^{\mu\nu}}$, the coset transforms as
\begin{equation}
gU\left(x;B_{Ai}\right)=\exp\left(ix^{\prime\mu}P_{\mu}\right)\exp\left(iB_{Ai}^{\prime}\left(x^{\prime}\right)X^{Ai}\right)h\left(B_{Ai}\left(x\right);g\right),
\end{equation}
where $h\left(B_{Ai};g\right)$ is of the form
\begin{equation}
h\left(B_{Ai};g\right)=e^{\frac{i}{2}\mu_{AB}\left(B_{Ai};g\right)T^{AB}}e^{\frac{i}{2}\mu_{ij}\left(B_{Ai};g\right)T^{ij}},\label{h}
\end{equation}
and from (\ref{pj alg}) and (\ref{pp alg}), we have 
\begin{equation}
\Omega(\omega)P_\mu\Omega^{-1}(\omega)=\Lambda(\Omega)^{\nu}_{\ \ \mu}P_\nu. \label{p trans}
\end{equation}
So, the coordinates transform in the usual way, 
\begin{equation}
x^{\prime\mu}=\Lambda_{\ \nu}^{\mu}x^{\nu}+a^{\mu}
\end{equation}
under a general Poincaré transformation.

The transformation of the NG fields is in general nonlinear and, as well as $\mu_{AB}$ and $\mu_{ij}$, can be calculated order by order in $\omega$ and $B_{Ai}$. To first order we have
\begin{eqnarray}
B^{\prime}_{Ai}(x^{\prime})&=&B_{Ai}+\omega_{Ai}+\omega_{ik}B_A^{\ \ k}+\omega_{AB}B^B_{\ \ i},\\
\mu_{AB}&=&\omega_{AB}-\omega_{Ai}B_B^{\ \ i},\\
\mu_{ij}&=&\omega_{ij}-\omega_{Ai}B^A_{\ \ j}.
\end{eqnarray}

When $g$ only involves elements of the invariant subgroups $H^{(1)}=SO(1,1)$ or $H^{(2)}=SO(2)$, the NG fields transform linearly. Namely,
\begin{equation}
B^{\prime}_{Ai}(x^{\prime})=h^{(1)B}_Ah^{(2)j}_iB_{Bj},\label{NG trans}
\end{equation}
with $h^{(1)}$ and $h^{(2)}$ belonging to $H^{(1)}$ and $H^{(2)}$, respectively. That is, only the coset elements transform the fields nonlinearly.

To construct invariant effective Lagrangians out of the NG fields, which is one of the main purposes of the present work, it is convenient to work with covariant objects instead of the NG fields directly. This is attained by considering the Maurer-Cartan form, according to our discussion in the last section. From (\ref{cos}), the expansion of the Maurer-Cartan form $U^{-1}dU$ in terms of the Poincaré algebra is given by
\begin{eqnarray}
U^{-1}dU & = & idx^{\mu}\Omega^{-1}\left(x\right)P_{\mu}\Omega\left(x\right)+\Omega^{-1}\left(x\right)d\Omega\left(x\right)\nonumber\\
 & = & idx^{\mu}\left(e_{\mu}^{\ A}P_{A}+e_{\mu}^{\ i}P_{i}+2D_{\mu}B_{Ai}X^{Ai}+\omega_{\mu AB}T^{AB}+\omega_{\mu ij}T^{ij}\right), \label{MC}
\end{eqnarray}
where we have defined the vielbeins $e_{\mu}^{\ A}$ and $e_{\mu}^{\ i}$ by the relation $\Omega^{-1}\left(x\right)P_{\mu}\Omega\left(x\right)=e_{\mu}^{\ A}P_A+e_{\mu}^{\ i}P_i$, according to (\ref{MC spacetime}). Using (\ref{p trans}), they can be written as
\begin{eqnarray}
e_{\mu}^{\ A}&=&\Lambda(\Omega)_\mu^{\ \ A}=\left(e^{B}\right)_\mu^{\ \ A},\label{viel h1}\\
e_{\mu}^{\ i}&=&\Lambda(\Omega)_\mu^{\ \ i}=\left(e^{B}\right)_\mu^{\ \ i}.\label{viel h2}
\end{eqnarray}
To get the last equality we have used the explicit form of the $X^{Ai}$ generators in the fundamental representation of the Lorentz group:
\begin{equation}
\left(X^{Ai}\right)_{\ \alpha}^{\rho}=-\frac{i}{2}\left(\eta^{A\rho}\delta_{\alpha}^{i}-\eta^{i\rho}\delta_{\alpha}^{A}\right).
\end{equation}

To obtain more compact expressions, we make progress with the notation and define a new kind of index using the lower case first half of the Latin alphabet, $a,\ b,\ c,\ldots$, which groups together both of the indices for $SO(1,1)$ and $SO(2)$ representations. Then, these new indices run from $0$ to $3$, but unlike the Greek indices, which have the same range and denote linear representations of $SO(1,3)$, the new ones transform nonlinearly under general $SO(1,3)$ transformations.

We also define antisymmetric NG fields $B_{ab}=\{B_{AB},\ B_{Ai},\ B_{ij}\}$, with $B_{AB}=B_{ij}\equiv 0$. So, $B_{ab}$ contains the same number of degrees of freedom as $B_{Ai}$. In the same way, we consider connections $\omega_{ab}=\{\omega_{AB},\ \omega_{Ai},\ \omega_{ij}\}$, with $\omega_{Ai}\equiv 0$, which includes both the connections appearing in (\ref{MC}).

With these definitions, we can rewrite (\ref{MC}) as
\begin{eqnarray}
U^{-1}dU = idx^{\mu}e_{\mu}^{\ a}\left(P_{a}+e^{\rho}_{\ a}D_{\rho}B_{bc}X^{bc}+e^{\rho}_{\ a}\omega_{\rho bc}T^{bc}\right), \label{compact MC}
\end{eqnarray}
where we have introduced the inverse vielbein $e^{\rho}_{\ a}$, according to the definitions (\ref{viel1}) and (\ref{viel2}). The quantity $D_aB_{bc}\equiv e^{\mu}_{\ a}D_{\mu}B_{bc}$ transforms in a fully covariant way under Poincaré transformations, as in the equation (\ref{cov der trans}), with $h$ $\in$ $SO(1,1)\otimes SO(2)$. In contrast, according to (\ref{con trans2}), $\omega_{abc}$ transforms covariantly in the first index, but as a connection in the last two ones.

Since the invariant subgroup is of the form of a product of groups, one can form independent invariants for each one of the groups. Therefore, we can define two independent metrics in the noncoordinate basis instead of one. From (\ref{viel h1}) and (\ref{viel h2}), $e^\mu_{\ a}$ is a Lorentz transformation, and then, it is also isometry of the Minkowski metric. So, we promptly have 
\begin{eqnarray}
\lambda_{AB}&=&e^\mu_{\ A}e^\mu_{\ B}\eta_{\mu\nu},\label{metric1}\\
\gamma_{ij}&=&e^\mu_{\ i}e^\mu_{\ j}\eta_{\mu\nu},\label{metric2}
\end{eqnarray}
with $\lambda$ and $\gamma$ being the same matrices already introduced in (\ref{lambda}) and (\ref{gamma}).
 
The explicit form of the covariant derivative and of the connection can be obtained at all orders in $B$. The calculation can be lengthy, but it is straightforward, and we just present them in the final form

\begin{eqnarray}
D_{a}B_{cd}X^{cd} & = & \frac{1}{2}\partial_{a}B_{bg}\left(B^{-1}\right)_{\ f}^{b}\sinh\left(\beta\right)^{fg;cd}X_{cd},\label{cov der}\\
\omega_{acd}T^{cd} & = & -\frac{1}{4}\partial_{a}B_{fg}\left(-2\delta_{c}^{f}\left(B^{-1}\right)_{\ d}^{g}+\left(B^{-1}\right)_{\ h}^{g}\cosh\left(\beta\right)_{\ \ ;cd}^{fh}\right)T^{cd},\label{con}
\end{eqnarray}
where we have defined
\begin{equation}
\beta_{ab;cd}\equiv\left(B_{ac}\eta_{bd}-B_{ad}\eta_{bc}\right),
\end{equation}
and $\left(B^{-1}\right)_{ab}$ by
\begin{equation}
\left(B^{-1}\right)_{ab}B^{bc}=\delta_a^c.
\end{equation}

At this point of the discussion we have all the ingredients to construct effective Lagrangians for the NG modes. Since the NG fields must enter only through the covariant derivative (\ref{cov der}), and this in turn is proportional to the usual derivative of the fields, an expansion of the effective Lagrangian in terms of increasing mass dimension operators is directly cast as an expansion in the number derivatives of the NG fields. When $S$-matrix elements involving these NG modes are calculated, these derivatives introduce factors of the NG characteristic energy of the process, and the expansion can be seen as an expansion in terms of this energy. As already emphasized, this is one of the main advantages of the nonlinear realization as compared to effective Lagrangians composed with linear representations of the broken symmetry group, where there is no such a convenient truncation of the Lagrangian for a given process \cite{weinberg2}.

However, among the known Standard-Model particles, there is no such a mode described by an antisymmetric $2$-tensor, and therefore we cannot directly relate scattering amplitudes involving only external NG particles to get phenomenological constraints on the effective Lagrangian coefficients. Nonetheless, the situation gets better when we consider the symmetry breaking mechanism taking place in a hidden sector of the Standard-Model and propagates to the other sectors through the coupling with the usual Standard-Model fields. We investigate this possibility in the following.

\subsection{Coupling to Matter}

As discussed in the review section, other fields that we generally call matter can be straightly introduced in the framework of the coset construction. For a field $\psi(x)$ belonging to a linear representation $\mathcal{D}$ of the invariant subgroup $SO(1,1)\otimes SO(2)$, its transformation under an arbitrary Poincaré group element is taken as $\mathcal{D}\left(h\left(B;g\right)\right)$, with $h$ corresponding to the compensating $H$-transformation defined in (\ref{h}). Furthermore, derivatives of the field can be introduced covariantly as in (\ref{cov mat}) with the connection (\ref{con}).

It is assumed that the heavy fields decouple at low-energies, according to the Appelquist-Carazzone theorem \cite{Appelquist}. Actually, the effect of these fields is encoded in all sort of operators allowed by symmetry that contributes to the effective Lagrangian. In this vein, we can restrict ourselves to the coupling of NG modes only with light fields.

In this work, we consider the possibility of the interaction of the photon field with the NG modes. This extended QED can modify much of the photon dynamics and should be confronted with the standard QED phenomenology. The masses of the other Standard-Model particles are sufficiently small compared to Planck Mass, which justifies a broader investigation of possible new effects brought by the coupling with NG modes. In spite of the interest, this extensive inspection will not be pursued here, since it lies beyond of the scope of the present work.

To consider small deviations from standard QED due to Lorentz-violating effects, it is still reasonable to consider $U(1)$ gauge invariance. To start with, we then consider a vector field $A_\mu(x)$ transforming linearly under the Lorentz group, whose dynamics is invariant under the gauge transformations
\begin{equation}
A_{\mu}^{\prime}=A_{\mu}+\partial_{\mu}\alpha.\label{gt linear}
\end{equation}
As is well known, this gauge invariance together with the equations of motion, reduces the number of dynamic degrees of freedom from $4$ to the $2$ photon polarizations.

Under transformations of the invariant subgroup $SO(1,1)\otimes SO(2)$, $A_\mu$ can be reduced to the direct sum of the linear representations $\left(A_0,\ A_1\right)$ and $\left(A_2,\ A_3\right)$. These, in turn, can be mapped to nonlinear representations $A_A$ and $A_i$ with the use of the inverse of the vielbeins (\ref{viel h1}) and (\ref{viel h2}). Thus, we have
\begin{equation}
A_{a}=e^{\mu}_{\ a}A_{\mu},
\end{equation}
with $A_a$ transforming under an arbitrary Poincaré transformation $g$ as 
\begin{equation}
A_{a}^{\prime}=\left(A^\prime_A,\ A^\prime_i\right)=\left(h^{(1)B}_AA_B,\ h^{(1)j}_iA_j\right),
\end{equation}
where, as in (\ref{NG trans}), $h^{(1)}$ and $h^{(2)}$ are $SO(1,1)$ and $SO(2)$ transformations, respectively, that can in general depend on the NG fields and the arbitrary $g$ transformation as in (\ref{h}).

Particularly, in the nonlinear representation the gauge transformation (\ref{gt linear}) reads
\begin{equation}
A_{a}^{\prime}=A_{a}+e^{\mu}_{\ a}\partial_{\mu}\alpha.\label{gauge nonlinear}
\end{equation}

Due to this invariance, the field $A_a$ always enter in the effective Lagrangian through the field strength $\tilde{F}_{ab}$ in the nonlinear representation, which relates to $F_{\mu\nu}$ in the linear case via $\tilde{F}_{ab}=e^{\mu}_{\ a}e^{\nu}_{\ b}F_{\mu\nu}$. Explicitly, we have
\begin{equation}
\tilde{F}_{ab} = \left(\left(D_{a}B_{b}^{\ d}-D_{b}B_{a}^{\ d}\right)A_{d}+\nabla_{a}A_{b}-\nabla_{b}A_{a}\right),\label{fs}
\end{equation}
which is invariant under the transformation (\ref{gauge nonlinear}), as can be readily shown.

\subsection{Effective action \label{eff ac}}

With the above discussion, we have all the elements to write down the most general effective Lagrangian with interacting NG modes and photons. We only need to combine NG covariant derivatives and field strengths (\ref{fs}) and contract them with the metrics (\ref{metric1}) and (\ref{metric2}) or, alternatively, with the Levi-Civita symbols
\begin{equation}
\epsilon_{AB}=\left(\begin{array}{cc}
0 & 1\\
-1 & 0
\end{array}\right)
\end{equation}
and
\begin{equation}
\epsilon_{ij}=\left(\begin{array}{cc}
0 & 1\\
-1 & 0
\end{array}\right),
\end{equation}
since, like $\lambda_{AB}$ and $\gamma_{ij}$, these are invariant tensors under the $SO(1,1)$ and $SO(2)$ transformations of the invariant subgroup.

All possible contractions of the covariant terms with metrics and Levi-Civita symbols result in the following action for NG modes interacting with photons up to two derivatives:
\begin{eqnarray}
S & = & \int d^{4}x\bigg(a_{1}\left(D_{C}B_{Ai}\right)\left(D^{C}B^{Ai}\right)+a_{2}\left(D_{j}B_{Ai}\right)\left(D^{j}B^{Ai}\right)+a_{3}\left(D_{A}B_{\ i}^{A}\right)\left(D_{C}B^{Ci}\right)+\nonumber\\
 &  & +a_{4}\left(D_{i}B^{Ai}\right)\left(D_{j}B_{A}^{\ \ j}\right)+a_{5}\left(D_{A}B_{\ \ i}^{C}\right)\left(D_{C}B^{Ai}\right)+a_{6}\left(D_{i}B^{Aj}\right)\left(D_{j}B_{A}^{\ \ i}\right)+\nonumber\\
 &  & +a_{7}\epsilon^{AB}\left(D_{A}B_{Bi}\right)\left(D_{C}B^{Ci}\right)+a_{8}\epsilon^{AB}\left(D_{A}B_{Ci}\right)\left(D^{C}B_{B}^{\ \ i}\right)+a_{9}\epsilon^{ij}\left(D_{i}B_{Bj}\right)\left(D_{l}B^{Bl}\right)+\nonumber\\
 &  & +a_{10}\epsilon^{ij}\left(D_{i}B_{Bl}\right)\left(D^{l}B_{\ \ j}^{B}\right)+a_{11}\epsilon^{AB}\epsilon^{ij}\left(D_{l}B_{Ai}\right)\left(D^{l}B_{Bj}\right)+a_{12}\epsilon^{AB}\epsilon^{ij}\left(D_{C}B_{Ai}\right)\left(D^{C}B_{Bj}\right)+\nonumber\\
 &  & +a_{13}\epsilon^{AB}\epsilon^{ij}\left(D_{A}B_{Ci}\right)\left(D_{B}B_{\ \ j}^{C}\right)+a_{14}\epsilon^{AB}\epsilon^{ij}\left(D_{i}B_{Al}\right)\left(D_{j}B_{B}^{\ \ l}\right)+a_{15}\nabla_A\left(D_{i}B^{Ai}\right)+\nonumber\\
& & +a_{16}\nabla_i\left(D_{A}B^{Ai}\right)+a_{17}\epsilon^{AB}\nabla_i\left(D_{A}B_B^{\ \ i}\right)+a_{18}\epsilon^{AB}\nabla_A\left(D_{i}B_{B}^{\ \ i}\right)+a_{19}\epsilon^{ij}\nabla_i\left(D_{A}B^{A}_{\ \ j}\right)+\nonumber\\
& & +a_{20}\epsilon^{ij}\nabla_A\left(D_{i}B^{A}_{\ \ j}\right)+a_{21}\epsilon^{AB}\epsilon^{ij}\nabla_A\left(D_{i}B_{Bj}\right)+a_{22}\epsilon^{AB}\epsilon^{ij}\nabla_i\left(D_{A}B_{Bj}\right)\nonumber\\
 &  & +b_{1}\tilde{F}_{AB}\tilde{F}^{AB}+b_{2}\tilde{F}_{ij}\tilde{F}^{ij}+b_{3}\tilde{F}_{Ai}\tilde{F}^{Ai}+b_{4}\epsilon^{AB}\epsilon^{ij}\tilde{F}_{AB}\tilde{F}_{ij}+b_{5}\epsilon^{AB}\tilde{F}_{Ai}\tilde{F}_{B}^{\ \ i}+b_{6}\epsilon^{ij}\tilde{F}_{Ai}\tilde{F}_{\ \ j}^{A}+\nonumber\\
 &  & +b_{7}\epsilon^{AC}\epsilon^{ij}\tilde{F}_{Ai}\tilde{F}_{Cj}+b_{8}\epsilon^{AB}\tilde{F}_{AB}+b_{9}\epsilon^{ij}\tilde{F}_{ij}\bigg).\label{eff action}
\end{eqnarray}
To get this effective action from the Lagrangian, we have integrated with the invariant measure (\ref{measure}). But, in the present case, $\text{det}\left(e_{\mu}^{\ a}\right)=1$, since $e_{\mu}^{\ a}$ is just a special Lorentz matrix.

It is beyond the scope of the present work to exhaust the analysis of the possible new effects brought by the plenty of terms displayed in the above action. Still, focusing on kinematical properties of the dynamical modes in this model and discussing some fundamental issues like stability and causality, we will be able to constrain some of the arbitrary parameters.

\section{Stability and causality: a general analysis\label{stability}}

Before delving into specific investigations, it is convenient to consider a simpler toy model to point some subtleties in the analysis of stability and causality in Lorentz-violating models. {\it En passant}, we will also clarify the role played by Lorentz symmetry in these models. Our discussion in this section follows closely others similar analysis made in Refs. \cite{alan-ralf,carroll,mukanov,null,arkani,superluminal}.

\subsection{Tachyon instabilities: exponentially growing modes}

To begin with, let us consider the simple model of a free scalar field in $3+1$ dimensions:
\begin{equation}
\mathcal{L}=\frac{1}{2}\left(\left(\partial_0\phi\right)^2-v^2\left(\partial_i\phi\right)^2\right),\label{l scalar}
\end{equation}
with $v^2\neq 1$.
Taking the light velocity as $c=1$, this Lagrangian is clearly non-invariant under Lorentz boosts. For simplicity, we will consider the behavior of the system under boosts in the $x$-direction:
\begin{eqnarray}
t^\prime&=&\frac{t-\beta x}{\sqrt{1-\beta^2}},\label{lt}\\
x^\prime&=&\frac{x-\beta t}{\sqrt{1-\beta^2}},\\
y^\prime&=&y,\\
z^\prime&=&z,\label{ltx}
\end{eqnarray}
with $\beta^2<1$.
To discuss the role of these transformations in the description of the dynamics of the model, let us use the tangent vector of rest observer (\ref{unit}) to rewrite the Lagrangian (\ref{l scalar}) as
\begin{equation}
\mathcal{L}_{n}=\frac{v^2}{2}\left(\partial_\mu\phi\partial^\mu\phi-\frac{\left(v^2-1\right)}{v^2}\left(n^\mu\partial_\mu\phi\right)^2\right).\label{eq4}
\end{equation}
Then, we can obviously verify the invariance $\mathcal{L}_{\Lambda n}\left(\Lambda x\right)=\mathcal{L}_{n}\left(x\right)$ under the coordinate changes (\ref{lt})-(\ref{ltx}).
By itself, this coordinate Lorentz invariance has no physical significance. It is referred  in Ref. \cite{sme} as {\it observer} Lorentz invariance, and it is a mere relabeling of the physical description being applicable to any Lagrangian, irrespective if the related dynamics respects or not special relativity. For the Lagrangian $\mathcal{L}$, for instance, the dynamics in two distinct coordinate systems may look like completely different from each other. To see this, take the equation of motion obtained from $\mathcal{L}_{n}$:
\begin{equation}
G^{\mu\nu}\partial_\mu\partial_\nu\phi(x)=0,\label{eq}
\end{equation}
where the effective metric $G^{\mu\nu}$ is given by
\begin{equation}
G^{\mu\nu}=\eta^{\mu\nu}-\frac{v^2-1}{v^2}n^\mu n^\nu,\label{ef metr}
\end{equation}
and the indices of equation (\ref{eq}) are raised and lowered with the Minkowski metric $\eta_{\mu\nu}$. For plane wave solutions, the frequency $\omega$ and wave vector $\vec{k}$ of the waves must satisfy the relation
\begin{equation}
G^{\prime\mu\nu}k^\prime_\mu k^\prime_\nu=0 \label{eq1}
\end{equation}
in an arbitrary coordinate system of the class defined by relations (\ref{lt})-(\ref{ltx}), with $k^\mu=\left(\omega,\vec{k}\right)$ and
\begin{equation}
G^{\prime \mu\nu}=\eta^{\mu\nu}-\frac{v^2-1}{v^2}\Lambda^\mu_{\ 0}\Lambda^\nu_{\ 0}.\label{eq2}
\end{equation}
Using this expression in (\ref{eq1}), we get the two solutions for the frequency in an arbitrary frame
\begin{equation}
\omega^\prime=\frac{\left(1-v^2\right)\beta k^\prime_x\pm \sqrt{\left(1-\beta^2\right)\left(v^2\left(1-\beta^2\right)k^{\prime 2}_x+\left(1-\beta^2v^2\right)k^2_\perp\right)}}{1-\beta^2v^2}.\label{fr}
\end{equation}

Let us now analyze the behavior of the modes for $v^2<1$ and $v^2>1$, in turn. From (\ref{eq}) and (\ref{ef metr}), we can see that $v$ correspond to the phase velocity of the modes in the rest frame defined by the coordinate system where $n^\mu$ has the form as in (\ref{unit}). Then, for subluminal velocities of these modes, $v^2<1$, the two solutions in (\ref{fr}) are always real for any $\left|\beta\right|$, as can be promptly verified. This means the evolution of the modes is stable according to arbitrary observers when these observers set up natural initial conditions in their reference frame.

The situation changes considerably when we consider waves with superluminal velocities, $v^2>1$. First we should mention that superluminal signals are a potential source of causality paradoxes. But letting this aside for awhile, let us continue investigating the issue of stability. In this case, we can see that the discriminant of the square root in (\ref{fr}) can become negative for $\beta>\frac{1}{v}$ and $k^2_\perp>\left(\frac{1-\beta^2}{\beta^2v^2-1}\right)k^{\prime 2}_x$, leading to exponentially growing and decaying modes. This should be interpreted as an instability in the dynamics of the modes with natural initial conditions posed by some observer defined by $\beta>\frac{1}{v}$. Let us clarify this point: it is clear that such an unstable solution cannot be excited in the rest frame, since we saw that no complex frequency appears in that frame, and by boosting a real frequency and real wave vector, we cannot get complex quantities. In other words, observer Lorentz transformations map solutions to solutions of equation (\ref{eq1}), but a given solution obtained from a sensible set of initial conditions for one observer may be a discarded solution by another observer, since it may not correspond to the evolution of natural initial conditions as judged by that observer. This is made evident when we boost unstable solutions for fast moving observers back to the rest frame. Then, we verify that modes satisfying $k^2_\perp>\left(\frac{1-\beta^2}{\beta^2v^2-1}\right)k^{\prime 2}_x$ correspond to complex wave vectors in the rest frame and cannot be superposed to construct sensible initial conditions in this frame.

This raises the question if it is possible at all for the moving observer to set up initial conditions that evolve unstably. In Ref. \cite{mukanov}, it is shown that the retarded Green function, defined directly in the moving frame, $G_R^{mf}\left(t^\prime,x^\prime\right)$, contains exponentially growing modes in momentum space and cannot be integrated to define a sensible Green function in coordinate space. That means this observer cannot adjust a source to send signals to his future --- growing $t^\prime$ --- and he cannot then pose the initial conditions that would evolve unstably. So, the correct Green function, which describes the response to any source in the moving frame, is the one got by boosting the retarded Green function $G_R^{rf}\left(t,x\right)$ defined in the rest frame. In the moving frame, this retarded Green function, $G_R^{rf}\left(t^\prime,x^\prime\right)$, is actually an admixture of the retarded and advanced Green functions, $G_R^{mf}\left(t^\prime,x^\prime\right)$ and $G_A^{mf}\left(t^\prime,x^\prime\right)$, for the primed coordinates. This reflects the fact that the time order of events, as seen from the moving frame, is reversed as compared to the rest frame and, as mentioned before, this could be a source of causality problems. However, this is not the case here as we now discuss.

\subsection{Causality}

In special relativity the presence of superluminal signals is threatening. It paves the way to build mechanisms to communicate with our own past and many causal paradoxes can be devised. The kind of reasoning to construct such inconsistencies can be summarized in the called ``tachyonic antitelephone" paradox. The situation is the following: consider an observer $S$ that sends a tachyonic signal with velocity $c_t>1$ to another observer $S^\prime$ that moves with velocity $v$ with respect to $S$. At the time the signal arrives at $S^\prime$, he sends a tachyon signal back to $S$. As discussed above, if $v>1/c_t$ the signal propagates backward in time in the rest frame of the observer $S$. Thus, the observer could, in principle, communicate with their own past. In other words, closed timelike curves could be constructed in this spacetime.

However, an underlying assumption in this kind of reasoning is the complete equivalence of all observers in agreement with special relativity. If, on the other hand, a reference frame for some reason can be singled out, then the existence of closed timelike curves can be avoided. For example: we can imagine a situation where the tachyon moves with a constant velocity in the preferred reference frame. Then, the signal can only moves forward in time in this frame and no closed timelike curve can be built. This seems to be exactly the case with the model we considered in this section, since once we assume $v^2>1$, equation (\ref{eq}) states that the modes move superluminally and always with the same velocity $v$ in the rest frame. This means that another observer can only send a tachyon moving to the future of the rest observer.

This discussion suggests that the presence of superluminal signals cannot be the sole reason for the existence of closed timelike curves. In fact, one can state precisely the conditions for the non-existence of closed timelike curves through the notion of {\it stable causality} \cite{wald}: {\it A spacetime $\left(\mathcal{M},g_{\mu\nu}\right)$ is stably causal if and only if there exists a differentiable function $f$ on $\mathcal{M}$, such that $\nabla_\mu f$ is a timelike vector field with respect to $g_{\mu\nu}$}. Here, $\nabla_\mu$ is a covariant derivative associated with the metric $g_{\mu\nu}$. The importance of this definition is that a stably causal spacetime possesses no closed timelike or closed null curve. In the case of an effective geometry the associated effective metric $G^{-1}_{\mu\nu}$ plays the role of the metric $g^{\mu\nu}$.

Now, considering the Minkowski time $t$ defined by $\partial_\mu t=n_\mu$ and using (\ref{ef metr}), we have
\begin{equation}
G^{\mu\nu}\partial_\mu t\partial_\nu t=\frac{1}{v^2}>0.
\end{equation}
Then $t$ can be taken as the function $f$ in the above theorem, and the effective geometry generated by $G^{-1}_{\mu\nu}$ renders a stably causal spacetime. This agrees with our previous analysis of the absence of closed timelike curves in spite of the presence of superluminal signals.

\subsection{Ghost instabilities}

Let us get back to the stability discussion. The model can also be unstable if there are {\it ghosts} in its spectrum. These are characterized by having negative kinetic energy. So, the free Hamiltonian is not bounded from below. Once these negative energy particle are set to interact with ordinary particles with positive free Hamiltonian, the instability is manifest. The quantum vacuum, for example, could spontaneously radiate positive and negative energy particles compatible with vanishing total energy.

Again, we will focus on the simple model (\ref{l scalar}) to get some intuition before going into the more complicated analysis of the model (\ref{eff action}). However, as stated above, we need to consider some interaction of the $\phi$ field with some other field to access possible ghost instabilities. A general analysis, with $\phi$ interacting with many different fields obeying Lorentz non-invariant dispersion relations, could be very intricate and would not be so much enlightening. Considering that in the context of the model (\ref{eff action}) we are only considering a NG mode interacting with the photon, and the current sensitivity for the  birefringent set of coefficients of the photon sector of the SME lies at $10^{-32}$, it is reasonable as a first check just allocate all Lorentz-violating effects on the NG sector and consider the relativistic dispersion relation for the photons. Then, for a matter of comparison it suffices just consider the coupling of $\phi$ with an extra field, we call $\chi$, moving along the light cone. Furthermore, still following the analogy with (\ref{l scalar}), we consider two kinds of vertices: one with one $\phi$ and two $\chi$'s, and the other with two $\phi$'s and two $\chi$'s.

Let us first assume the $\phi$ modes move subluminally in the rest frame --- $v^2<1$ in (\ref{l scalar}). Then, taking $k_\perp=0$ in (\ref{fr}), we get
\begin{equation}
\omega^\prime_+=\frac{\left(v+\beta\right) k^\prime_x}{1+\beta v},\ \ \ \omega^\prime_-=\frac{\left(\beta-v\right) k^\prime_x}{1-\beta v}.\label{eq3}
\end{equation}

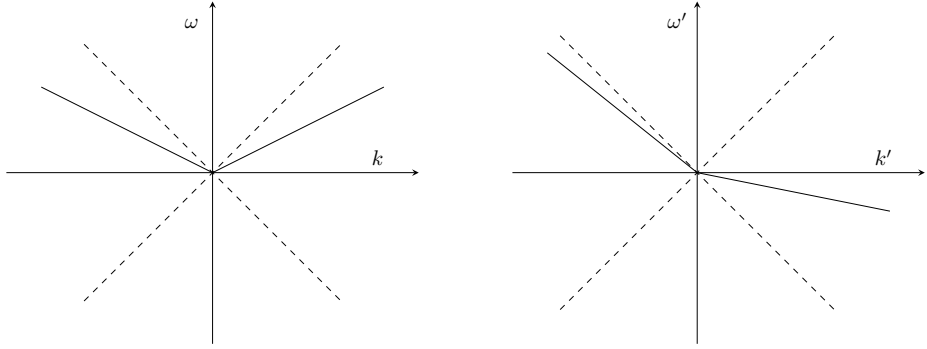
\begin{figure}
\begin{tikzpicture}[scale=0.8]
\begin{axis}[ticks=none,axis x line=center,axis y line=center,xlabel=$k$,ylabel=$\omega$,axis equal,x label style={at={(axis description cs:0.9,0.5)},anchor=south},y label style={at={(axis description cs:0.45,0.9)},anchor=south},ymin=-4,ymax=4]
\addplot[domain=-3:3,style=dashed]{x};
\addplot[domain=-3:3,style=dashed]{-x};
\addplot[domain=-4:0]{-0.5*x};
\addplot[domain=0:4]{0.5*x};
\end{axis}
\end{tikzpicture}
\hspace{1cm}
\begin{tikzpicture}[scale=0.8]
\begin{axis}[ticks=none,axis x line=center,axis y line=center,xlabel=$k^\prime$,ylabel=$\omega^\prime$,axis equal,x label style={at={(axis description cs:0.9,0.5)},anchor=south},y label style={at={(axis description cs:0.4,0.9)},anchor=south},ymin=-4,ymax=4]
\addplot[domain=-3.2:3.2,style=dashed]{x};
\addplot[domain=-3.2:3.2,style=dashed]{-x};
\addplot[domain=-3.5:0]{-0.8*x};
\addplot[domain=0:4.5]{-0.2*x};
\end{axis}
\end{tikzpicture}
\caption{$k_\perp =0$ section of the dispersion relations (undotted lines) for a massless particle satisfying eq. (\ref{eq3}) when $v^2<1$: in the rest frame ($\beta=0$) (left), and in a highly boosted frame ($\beta<-\left|v\right|$) (right).}\label{fig1}
\end{figure}

For $\left|\beta\right|>v$, one of the branches of the dispersion relation dives below the $\omega=0$ axis (fig. \ref{fig1}), exhibiting dangerous negative free energy. We can also investigate this issue accessing the free Hamiltonian of the model (\ref{l scalar}):
\begin{equation}
H_{\phi}=\frac{1}{2v^2G^{00}}\left(\Pi_{\phi}-G^{0i}\partial_i\phi\right)^2-\frac{v^2}{2}G^{ij}\partial_i\phi\partial_j\phi,\label{ham}
\end{equation}
with the canonical momentum given by
\begin{equation}
\Pi_{\phi}=v^2G^{0\mu}\partial_\mu\phi.
\end{equation}

For a boost in the $x$ direction, we get from (\ref{eq2})
\begin{eqnarray}
G^{00}&=&\frac{1-\beta^2 v^2}{v^2\left(1-\beta^2\right)},\\
G^{0i}&=&\frac{\beta\left(v^2-1\right)}{\left(1-\beta^2\right)v^2}\delta^{1i},\\
G^{11}&=&\frac{1-\beta v^2}{v^2\left(1-\beta^2\right)},\\
G^{ij}&=&-1,\ \ \ i,\ j\neq 1.
\end{eqnarray}
Then, for $v^2<1$ we notice $G^{00}>0$ and $G^{22}=G^{22}<0$, but $G^{11}>0$ for $\left|\beta\right|>v$, and the Hamiltonian (\ref{ham}) is not bounded neither from above or from below.

The above problems with the energy of the particles (\ref{eq3}) and with the Hamiltonian (\ref{ham}) for a fast moving observer are disturbing, since in the rest frame the model is completely fine. In fact, we can show that this is just an apparent instability. This is the case because the processes would trigger the instabilities are kinematically forbidden in the fast moving frame: the two vertices $\phi\chi^2$ and $\phi^2\chi^2$ could give rise to the vacuum instability due to the decays {\it vacuum} $\rightarrow$ $2\chi+\phi$ and {\it vacuum} $\rightarrow$ $2\chi+2\phi$, respectively, where one of the $\phi$-particles in each process would have a negative energy. However, by energy-momentum conservation, the four-momentum of one of the $\phi$-particle in both processes should lie in the past light-cone to the processes be possible, but, from the fig. \ref{fig1}, this will never be the case.

\begin{figure}
\begin{tikzpicture}[scale=0.8]
\begin{axis}[ticks=none,axis x line=center,axis y line=center,xlabel=$k$,ylabel=$\omega$,axis equal,x label style={at={(axis description cs:0.9,0.5)},anchor=south},y label style={at={(axis description cs:0.45,0.9)},anchor=south},ymin=-5.5,ymax=5.5]
\addplot[domain=-4:4,style=dashed]{x};
\addplot[domain=-4:4,style=dashed]{-x};
\addplot[domain=-2.8:0]{-1.8*x};
\addplot[domain=0:2.8]{1.8*x};
\end{axis}
\end{tikzpicture}
\hspace{1cm}
\begin{tikzpicture}[scale=0.8]
\begin{axis}[ticks=none,axis x line=center,axis y line=center,xlabel=$k^\prime$,ylabel=$\omega^\prime$,axis equal,x label style={at={(axis description cs:0.9,0.5)},anchor=south},y label style={at={(axis description cs:0.45,0.9)},anchor=south},ymin=-6,ymax=6]
\addplot[domain=-4.5:4.5,style=dashed]{x};
\addplot[domain=-5:5,style=dashed]{-x};
\addplot[domain=0:4]{1.25*x};
\addplot[domain=0:1.2,style=dotted]{5*x};
\addplot[domain=-1.2:0]{5*x};
\end{axis}
\end{tikzpicture}
\caption{$k_\perp =0$ section of the dispersion relations (undotted lines) for a massless particle satisfying eq. (\ref{eq3}) when $v^2>1$: in the rest frame ($\beta=0$) (left), and in a highly boosted frame ($\beta<-\frac{1}{\left|v\right|}$) (right).}\label{fig2}
\end{figure}
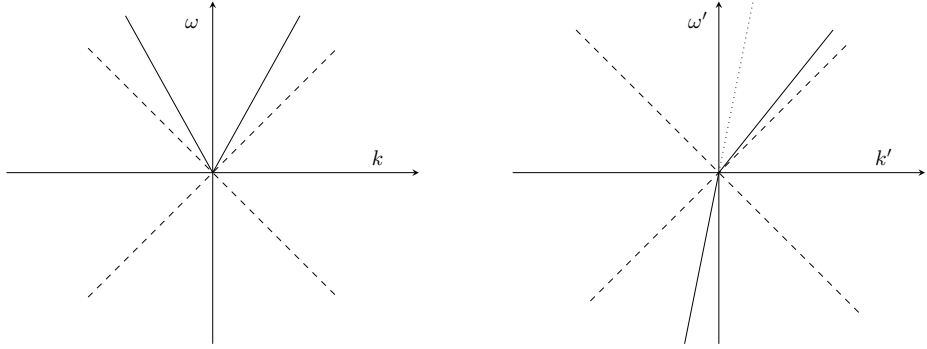

\begin{figure}
\begin{tikzpicture}[scale=0.8]
\begin{axis}[ticks=none,axis x line=center,axis y line=center,xlabel=$x$,ylabel=$t$,axis equal,x label style={at={(axis description cs:0.9,0.5)},anchor=south},y label style={at={(axis description cs:0.45,0.9)},anchor=south},ymin=-4,ymax=4]
\addplot[domain=-3:3,style=dashed]{x};
\addplot[domain=-3:3,style=dashed]{-x};
\addplot[domain=-4:0]{-0.5*x};
\addplot[domain=0:4]{0.5*x};
\end{axis}
\end{tikzpicture}
\hspace{1cm}
\begin{tikzpicture}[scale=0.8]
\begin{axis}[ticks=none,axis x line=center,axis y line=center,xlabel=$x^\prime$,ylabel=$t^\prime$,axis equal,x label style={at={(axis description cs:0.9,0.5)},anchor=south},y label style={at={(axis description cs:0.45,0.9)},anchor=south},ymin=-4,ymax=4]
\addplot[domain=-3:3,style=dashed]{x};
\addplot[domain=-3:3,style=dashed]{-x};
\addplot[domain=-3.8:0]{0.2*x};
\addplot[domain=0:3.3]{0.8*x};
\end{axis}
\end{tikzpicture}
\caption{$y=z=0$ section of the light-cone (undotted lines) for a massless superluminal particle: in the rest frame ($\beta=0$) (left), and in a highly boosted frame ($\beta<-\frac{1}{\left|v\right|}$) (right).}\label{fig3}
\end{figure}
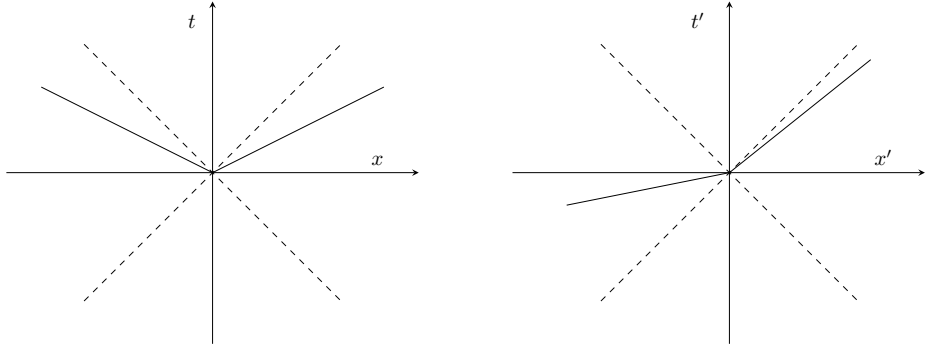

The situation seems to be more problematic for superluminal $\phi$-particles. For $v^2>1$ and $\left|\beta\right|>1/v$, the effective metric component $G^{00}$ becomes negative, and we again have an unbounded Hamiltonian. The two branches of $\omega$ transform as in fig. \ref{fig2}. The signs of $\omega$ and $k$ must be reversed after the transformation because we are defining particles as moving forward in the coordinate time: the $\phi$-particles move superluminally, and a positive time interval $\Delta t$ in the rest frame corresponds to negative $\Delta t^\prime$ in the fast moving frame. Then, we reinterpret a particle with positive energy and momentum moving backward in time as a negative energy and momentum particle propagating forward in time (fig. \ref{fig3}). Unlike the subluminal case, a negative energy $\phi$-particle can abide the past light cone in the moving frame, and processes $\phi\rightarrow 2\chi$ and $\phi\rightarrow \phi+2\chi$ in the rest frame would be seen as the vacuum decays {\it vacuum} $\rightarrow$ $2\chi+\phi$ and {\it vacuum} $\rightarrow$ $2\chi+2\phi$, respectively, corresponding to a true vacuum instability.

As in the analysis of the tachyon instabilities, this result seems to be paradoxical, since the system is perfectly stable in the rest frame. The solution to this puzzle is also similar to the tachyon case. There, the causal properties, as measured directly in the fast moving frame, pointed to unstable evolution of deemed natural initial conditions in that frame. However, we argued that these natural initial conditions could not be built by the moving observer, and the correct initial set-up is the one obtained by boosting natural initial conditions, posed in the rest frame, to the moving one. Concerning the ghost instabilities, the situation is analogous. In fact, the two observers define inequivalent vacua, whose choices are closely related to their notions of causality \cite{null}. In canonical quantization, for example, one need to define a space-like hypersurface upon which the quantum operators satisfy commutation relations, and this implicitly picks a vacuum where one creates particle propagating forwarding in the time orthogonal to the hypersurface. Since in the presence of superluminal signal the causal cone is broader than the light cone, the hypersurface chosen to define the quantization of the system should be spacelike also with respect to effective metric $G^{\mu\nu}$, defined in (\ref{ef metr}). Now, this is true for the hypersurface $t=0$ of the rest observer, but it is not for the choice $t^\prime=0$ of the fast moving frame. Then, the $t^\prime=0$ hypersurface is inside the causal cone and does not constitute a sensible choice of Cauchy initial data surface. This implies that the natural vacuum is the one chosen by the rest observer and, once this choice is made, no instability will show up.

\subsection{Spontaneous vs Explicit symmetry breaking}

All the analysis of this section refers to the model (\ref{l scalar}) or, equivalently, to the Lorentz coordinate invariant version (\ref{eq4}). In spite of the observer invariance, the system is not invariant under {\it particle} Lorentz transformations, which are defined as Lorentz transformations of the dynamical fields while keeping the background vector $n^\mu$ fixed. This means the Lorentz symmetry is explicitly broken in the model. As previously stressed, the observer Lorentz invariance is then forced by hand and, by itself, does not have any special significance. However, this invariance can be extremely convenient if we fairly suppose the matter and predominant interactions that make up the observer instruments respect Lorentz invariance. Furthermore, we could think the model (\ref{eq4}) as the quadratic part of an invariant system with the Lorentz symmetry spontaneously broken by considering perturbation around a non-invariant background. For example, we take the effective Lorentz-invariant Lagrangian
\begin{equation}
\mathcal{L}=\frac{1}{2}\partial_\mu\sigma\partial^\mu\sigma+\frac{b}{\Lambda^4}\left(\partial_\mu\sigma\partial^\mu\sigma\right)^2,\label{eq6}
\end{equation}
with $\Lambda$ being some mass scale. Higher dimension operators are suppressed by higher powers of the cutoff $\Lambda$. If we consider perturbations $\partial_\mu\phi=\partial_\mu\sigma-C_\mu$ around a nontrivial constant background $\partial_\mu\sigma_0=C_\mu$, we get the Lagrangian 
\begin{equation}
\mathcal{L}=\frac{v^2}{2}\partial_\mu\phi\partial^\mu\phi+\frac{v^2-1}{2}n^\mu n^\nu\partial_\mu\phi\partial_\nu\phi+\frac{4b}{\Lambda^4}C^\mu\partial_\mu\phi\partial_\nu\phi\partial^\nu\phi+\frac{b}{\Lambda^4}\left(\partial_\mu\phi\partial^\mu\phi\right)^2,\label{eq5}
\end{equation}
with $\frac{2b}{\Lambda^4}C^\mu C^\nu=\frac{(v^2-1)}{2}n^\mu n^\nu$, and we have discarded constant and total derivative terms.

The quadratic part of the Lagrangian (\ref{eq5}), which governs the dynamics of the free modes in the model, is identical to the model (\ref{eq4}) we have considered in our discussions. But here, the observer Lorentz invariance is a consequence of the Lorentz symmetry of the underlying model (\ref{eq6}), and does not need to be imposed by hand as in (\ref{eq4}). This is only a difference in principle, not in effect, and therefore the same conclusions we arrived by investigating causality and stability issues of the system (\ref{eq4}) are applied to the quadratic part of (\ref{eq5}). However, it is worth mentioning that the inherited observer Lorentz invariance of the model (\ref{eq5}) is not equivalent to the Lorentz invariance of (\ref{eq6}). The latter has much stronger dynamic implications and leads to invariance of the full model (\ref{eq5}) under particle Lorentz transformations realized nonlinearly on the field $\phi$. This, in turn, gives the propagating mode the status of a NG boson.

The path we took to the construction of the action (\ref{eff action}) was slightly different than the one to the action (\ref{eq5}). Since the microscopic model that would generate the effective action (\ref{eff action}) is not known, we focused from the onset on the breaking symmetry pattern and the low-energy degrees of freedom, ignoring any reference to the linear representations of the Lorentz group. Thence, we did not get an observer invariant Lagrangian. As we will show explicitly, but can be easily guessed, the dispersion relations of the free propagating modes resulting from the action (\ref{eff action}) are of the general form
\begin{equation}
p_Ap^A+\alpha p_ip^i+\beta=0,\label{eq7}
\end{equation}
with $\alpha$ and $\beta$ being some constant combinations of the Lagrangian parameters. This index structure refers to the representations of the invariant subgroups $SO(1,1)$ and $SO(2)$ of the background $\bar{\theta}_{\mu\nu}$. However, we recall that this expression is also invariant under general particle Lorentz transformations, since these act on the indices $A$ and $i$ as nonlinear $H$ transformations. The reference frame was fixed once and for all when we fixed the form of the $\bar{\theta}_{\mu\nu}$ in (\ref{theta}). In that coordinate system, the observer four-velocity has the form (\ref{unit}), and then it corresponds to the observer rest frame. So, a observer Lorentz transformation changes the simple form of $\bar{\theta}_{\mu\nu}$ and $n^\mu$, and to perform it on the Lagrangian (\ref{eff action}) we can just consider the $H$ representations as transforming linearly under arbitrary Lorentz transformations. In this way, we can apply to the dispersion relations like (\ref{eq7}) the same reasoning we applied in the stability and causality analysis of this section. In particular, they are subject to the same potential problems when $\alpha^2\neq 1$, and the solutions to all the apparent contradictions are also the same: once the system does not have any pathology in the rest frame, it will not have in any other frame.

Our assumption in this work is that the violation of Lorentz symmetry is spontaneous, and then we expect there would be some Lorentz-invariant UV completion of the theory. As we have seen, the effective theory can be consistent even in the presence of superluminal velocities for the modes. Related to this point, there is the interesting discussion if such a kind of effective model could be embedded in a local UV complete and Poincaré-invariant theory \cite{arkani}, since a renormalizable Poincaré-invariant theory is endowed with the microcausality condition and the consequent subluminal propagation of any signal. For this reason, we will also consider the constraint $v^2\leq 1$ in the forthcoming analysis.

\section{Linear stability and Causality of the low-energy effective model\label{causality}}

Following our discussion of the foregoing section, it is enough to investigate the stability and causality properties of the action (\ref{eff action}) in the rest frame, where $\bar{\theta}_{\mu\nu}$ has the coordinate form (\ref{theta}). Causality is accessed by means of the analysis of the effective metric as in (\ref{eq}), which has that same form for the dynamical modes of (\ref{eff action}), as we show in the Appendix through the analysis of their dispersion relations. Concerning the linear stability, we need to impose constraints in the free parameters of the action to inhibit the appearance of ghosts and tachyons in the quadratic Lagrangian.

To pursue this analysis, we first get the quadratic Lagrangian from (\ref{eff action}) by considering the covariant derivatives at first order in the fields. Then, we have
\begin{eqnarray}
\mathcal{L}_2 & = & \bigg(a_{1}\left(\partial_{C}B_{Ai}\right)\left(\partial^{C}B^{Ai}\right)+a_{2}\left(\partial_{j}B_{Ai}\right)\left(\partial^{j}B^{Ai}\right)+\left(a_{3}+a_5\right)\left(\partial_{A}B^{A}_{\ \ i}\right)\left(\partial_{C}B^{Ci}\right)+\nonumber\\
 &  & +\left(a_{4}+a_6\right)\left(\partial_{i}B^{Ai}\right)\left(\partial_{j}B_{A}^{\ \ j}\right)+\left(a_{7}+a_8\right)\epsilon^{AB}\left(\partial_{A}B_{Bi}\right)\left(\partial_{C}B^{Ci}\right)+\nonumber\\
 &  & +\left(a_{9}+a_{10}\right)\epsilon^{ij}\left(\partial_{i}B_{Bj}\right)\left(\partial_{l}B^{Bl}\right)+a_{11}\epsilon^{AB}\epsilon^{ij}\left(\partial_{l}B_{Ai}\right)\left(\partial^{l}B_{Bj}\right)+\nonumber\\
 &  & +a_{12}\epsilon^{AB}\epsilon^{ij}\left(\partial_{C}B_{Ai}\right)\left(\partial^{C}B_{Bj}\right)+b_{1}F_{AB}F^{AB}+b_{2}F_{ij}F^{ij}+b_{3}F_{Ai}F^{Ai}+\nonumber\\
 &  & +\left(b_{4}+\frac{1}{2}b_7\right)\epsilon^{AB}\epsilon^{ij}F_{AB}F_{ij}+b_{5}\epsilon^{AB}F_{Ai}F_{B}^{\ \ i}+b_{6}\epsilon^{ij}F_{Ai}F^{A}_{\ \ j}\nonumber\\
 &  & -2b_{8}\epsilon^{AB}B^i_{\ A}F_{iB}-2b_{9}\epsilon^{ij}B^A_{\ \ j}F_{Aj}\bigg),\label{eq10}
\end{eqnarray} 
where $F_{Ai}=\delta^\mu_A\delta^\nu_iF_{\mu\nu}$, and we have used integration by parts and discarded the terms that do not contribute at the quadratic level.

A ghost can be identified directly from the Lagrangian (\ref{eq10}) as a mode with a negative kinetic term. A nice way to identify the potential ghost modes from the degrees of freedom of an arbitrary tensor is to decompose this tensor into transverse and longitudinal pieces with respect to the derivative operator. For the NG modes of our model, we consider the following decomposition
\begin{equation}
B_{Ai}=\frac{1}{g^2}\left(B^\perp_{Ai}+\frac{1}{m^2}\partial_A\partial_i\pi\right),\label{eq8}
\end{equation}
where the field $B^\perp_{Ai}$ satisfy
\begin{equation}
\partial_A\partial_iB^{\perp Ai}=0.\label{eq9}
\end{equation}
The dimensionless parameter $g$ was introduced to control the scaling of the original field $B_{Ai}$ as we change the mass parameter $m$. This is a small mass parameter scale different from the cutoff scale $\Lambda$ of the effective theory. The two masses are in fact related by $m=g\Lambda$, and we are considering $g$ small. The convenience of the introduction of these parameters will become clear soon.

Plugging the decomposition (\ref{eq8}) back into the quadratic Lagrangian (\ref{eq10}), we get potentially higher derivatives whenever longitudinal degrees of freedom show up. After the substitution, we still get mixing terms between the transverse and longitudinal sectors. Taking the limit $g\rightarrow 0$, while keeping $\Lambda$ fixed, we find a window $m\ll E\ll\Lambda$, where the mixing terms are irrelevant compared to the unmixed ones and can be safely discarded. Now, we have the transverse and longitudinal sectors completely decoupled, with the dynamics for the latter having higher time or spatial derivatives, which leads to ghost instabilities. Strictly speaking only higher time derivatives are problematic, but since only symmetry constrains the form of the effective Lagrangian, if we allow terms like $\partial_A\partial_i\pi\partial^A\partial^i\pi$, we should also consider terms of the form $\partial_A\partial^A\pi\partial_B\partial^B\pi$, since both have the same scaling dimension.

The unhealthy terms can be avoided if we demand the Lagrangian to be gauge invariant under the symmetry 
\begin{equation}
B^{\prime}_{Ai}=B_{Ai}+\partial_A\partial_i\pi.\label{eq13}
\end{equation}
At the quadratic level this implies we need to impose the following constraints on the parameters
\begin{eqnarray}
\left(a_{3}+a_{5}\right)&=&-a_1,\label{eq11}\\
\left(a_{4}+a_{6}\right)&=&-a_2,\\
\left(a_{7}+a_{8}\right)&=&\left(a_{9}+a_{10}\right)=b_{8}=b_{9}=0.\label{eq12}
\end{eqnarray}

We show in the Appendix that the quadratic Lagrangian (\ref{eq10}) subjected to these constraints describes a NG mode and two photon polarizations. All the modes have dispersion relations of the form $p_Ap^A+\alpha p_ip^i=0$, as expected, corresponding to massless particles propagating anisotropically for general parameters. The phase velocity along the $x^2$ and $x^3$ axes is given by $\sqrt{\alpha}$. The requirements of subluminality and real frequencies leads to the condition $0\leq\alpha\leq 1$.

For the NG sector, we first see from (\ref{a9}) that we need to impose
\begin{equation}
a_1<0, \ \ \ a_2<0
\end{equation}
to avoid ghost and tachyon instabilities. From (\ref{a10}), the subluminality condition implies
\begin{equation}
\left|a_2\right|<\left|a_1\right|.
\end{equation}

The photon Lagrangian in (\ref{a1}) is a particular case of the photon sector considered in the SME framework. This could be made explicit by rewriting the Lagrangian as a sum of the standard Lorentz-invariant QED operators plus Lorentz-violating deviations. Then, we could identify the parameters $b_1,\ldots,b_7$ as a subset of the coefficients that appear in the Lorentz-violating operator $\left(k_F\right)_{\mu\nu\rho\sigma}F^{\mu\nu}F^{\rho\sigma}$ of the SME photon sector \cite{sme}. Cavity experiments and astrophysical observations put stringent bounds on the size of these coefficients \cite{Russell}. The components associated to birefringent effects --- $u^{+}\neq u^-$ in (\ref{a20}) --- are roughly bounded at $\mathcal{O}\left(10^{-32}\right)$, while those that only cause non-birefringent Lorentz deviations are bounded at $\mathcal{O}\left(10^{-17}\right)$.

An interesting question that should be addressed concerns the naturalness of spontaneously Lorentz broken models. The smallness of the Lorentz-violating coefficients in the Standard-Model sector strongly suggests the exactness of Lorentz symmetry not only for the dynamics of the Standard-Model fields, but also for any other field that could interact with them to avoid radiative induced Lorentz-violating effects. Without an protective mechanism, like the existence of some extra symmetry, which could postpone the appearance of Lorentz-violating operators to highly suppressed nonrenormalizable corrections, the UV complete theory should be awkwardly fine-tuned to allow such unnatural small deviations. In spite of the crucial importance of this matter for the full consistency of Lorentz-violating scenarios, we will just assume that such a mechanism exists, and we will investigate the consequences thenceforth.

Though tiny, the Lorentz-violating effects in the photon sector must be subjected to same subluminality and stability conditions we have been discussed. Then, from (\ref{a21}) we impose
\begin{equation}
rs>0,\ \ st>0,\ \ r^2-4st\geq 0,\ \ s\left(s+r+t\right)\geq 0,\label{eq40}
\end{equation}
which ensure $0\leq u^{\pm}\leq 1$.

Identifying the parameters $b_1,\ldots,b_7$ with the $k_F$ coefficients, as described above, and using the bounds listed in \cite{Russell}, we have
\begin{eqnarray}
b_1,\ b_2,\ b_3&=&-\frac{1}{4}+\mathcal{O}\left(10^{-\alpha}\right)\\
b_4+\frac{1}{2}b_7,\ b_5,\ b_6&\sim&\mathcal{O}\left(10^{-\alpha}\right),
\end{eqnarray}
with $\alpha$ varying from $13$ to $32$ roughly. Using (\ref{a19}) and (\ref{a22}), we then conclude that the two first conditions in (\ref{eq40}) are by far satisfied, since they could differ from $10^{-2}$ in one part in $10^{-2\alpha}$. For Lorentz-invariant QED, the last two conditions exactly vanish, and, in our Lorentz-violating scenario, they could differ from zero in one part in $10^{-2\alpha}$. However, it is interesting to notice that even a tiny Lorentz violation signal should give only positive contributions in the last two conditions of (\ref{eq40}).

We end our discussion with a few considerations about the nonlinear generalization of the gauge symmetry (\ref{eq13}) and its implication to the properties of the NG-photon interactions.

A natural way to generalize the symmetry given by (\ref{eq13}) is to introduce the NG fields in the following way: we start with the field strength $K_{\mu\nu\rho}=\partial_\mu\Gamma_{\nu\rho}+\partial_\rho\Gamma_{\mu\nu}+\partial_\nu\Gamma_{\rho\mu}$, with $\Gamma_{\mu\nu}$ being a antisymmetric field and then we make the replacement $\Gamma_{\mu\nu} \longrightarrow\Omega_{\mu\nu}$, where $\Omega_{\mu\nu}$ is the vector representation of $\Omega(B_{Ai})$ defined in (\ref{cos}). This is analogous to the procedure of going from a linear to a nonlinear sigma model when a symmetry is spontaneously broken. If we construct the effective Lagrangian only with $K_{\mu\nu\rho}$, $F_{\rho\sigma}$, and Lorentz metrics, we will get after the replacement a model invariant under $A_{\mu}^\prime=A_{\mu}+\partial_{\mu}\alpha$ and $\Omega_{\mu\nu}^\prime=\Omega_{\mu\nu}+\partial_\mu\partial_\nu\lambda$. Since $\Omega_{\mu\nu}$ is constrained to satisfy $\Omega_{AB}=\Omega_{ij}=0$, the last symmetry is the generalization of (\ref{eq13}). Furthermore, due to these constraints, $\Omega_{\mu\nu}$ is not a linear representation of the Lorentz group, and Lorentz symmetry is then spontaneously broken. Effectively, the only non-vanishing components of $K_{\mu\nu\rho}$ are $K_{ABi}$ and $K_{iAj}$. At the linear order, these coincide with $G_{ABi}$ and $G_{iAj}$ appearing in (\ref{a1}). But, following this construction, we cannot have completely arbitrary coefficients, as the $a$'s and $b$'s in (\ref{a1}), since to be independently Lorentz invariant these terms need to be contracted with the vielbeins (\ref{viel h1}) and (\ref{viel h2}), which are  not gauge invariant. In fact, as stated before, the gauge invariance requires we contract the $K_{\mu\nu\rho}$ and $F_{\rho\sigma}$ field strengths only with Lorentz metrics, which would implies $a_1=a_2$, $b_1=b_2=1/2b_3$, and $b_4+1/2b_7=b_5=b_6=0$. It is even tempting to relax the constraints of $\Omega_{\mu\nu}$ and keep $\Gamma_{\mu\nu}$ as describing the Golstone dynamics, since now we have a larger gauge symmetry $\Gamma_{\mu\nu}^\prime=\Gamma_{\mu\nu}+\partial_\mu\xi_\nu-\partial_\nu\xi_\mu$, and the conditions $\Gamma_{AB}=\Gamma_{ij}=0$ can be seen as possible gauge fixing conditions. The issue if Lorentz violation would be completely unobservable in this case is interesting, but still needs a formal proof.

\section{Summary and Conclusions\label{sum}}

In this article we considered the problem of the spontaneous breaking of Lorentz symmetry by the vacuum condensation of an antisymmetric $2$-tensor. Our aim was to describe the low-energy dynamics of the NG bosons interacting with photons. Using the coset framework for the construction of effective actions, we were able to write down the most general effective Lagrangian compatible with broken global symmetry pattern {\it Lorentz} $\longrightarrow$ {\it vacuum invariant subgroup} and the local $U(1)$ gauge invariance of QED. However, we still allowed broken-Lorentz operators in the QED sector in such a way that the quadratic low-energy photon Lagrangian is a subset of the SME photon sector. This identification automatically sets stringent bounds in the photon Lagrangian parameters to be consistent with the known phenomenology.

The requirement that the effective action is still Lorentz invariant, though in a non-linear way, imposes non-trivial restrictions on the form of the interactions of the NG modes. We considered terms in the effective action up to two derivatives, which contain highly non-linear NG self-interactions as well as NG-photon interactions. In spite of the non-linearities, these terms are still within the regime of validity of the effective model and their contributions can be consistently considered.

As discussed in Sec. \ref{stability}, the analysis of stability and causality of models with broken Lorentz symmetry is tricky. Within the context of spontaneous symmetry breaking, the problem still gets extra subtleties. Therefore, we have made a discussion of the main difficulties that may arise in this scenario by considering a simple enough toy model for a scalar field that exhibits all the potential problems we find in the more complex model that we investigate in this article. By pursuing this analysis, we obtained the conditions that an effective model, viewed as a low-energy limit of a local Lorentz-invariant UV complete theory, must satisfy to claim causality and linear stability. We then applied these conditions to our specific model in Sec. \ref{causality}, and we concluded that, for general parameters of the initial effective Lagrangian, the NG sector does not satisfy the required conditions. We then proposed an extra symmetry of the same kind of the Kalb-Ramond field to protect the model against the appearance of the longitudinal unstable modes in higher order terms. The final form of the effective Lagrangian propagates just one NG mode interacting with the two photon polarizations. This Lagrangian can be the starting point for future phenomenological investigations. The consistency of the effective Lagrangian considering higher order field interactions still needs careful analysis. Particularly, if observable Lorentz-violating dynamical effects give in fact contributions.

\section*{Acknowledgments}

It is a pleasure to thank V. Alan Kostelecký for the useful suggestions. I also would like to thanks Riccardo Penco, Claudia de Rham, and Sergei Dubovsky for the correspondence. I am grateful to Pedro R. S. Gomes for his constant help during all the preparation of this article. This work has been supported by CAPES (Coordena\c{c}\~{a}o de Aperfei\c{c}oamento de Pessoal de N\'{\i}vel Superior-Brazil).

\appendix

\section*{Appendix: Dispersion relations \label{appendix A}}

In this Appendix, we will consider the free equations of motion obtained from the quadratic Lagrangian (\ref{eq10}), subjected to the constraints (\ref{eq11})-(\ref{eq12}), to obtain the dispersion relations for the propagating modes.

Using (\ref{eq11})-(\ref{eq12}) in (\ref{eq10}), we obtain
\begin{eqnarray}
\mathcal{L}_2 & = & 2a_{1}G_{ABi}G^{ABi}+2a_{2}G_{iAj}G^{iAj}+b_{1}F_{AB}F^{AB}+b_{2}F_{ij}F^{ij}+b_{3}F_{Ai}F^{Ai}+\nonumber\\
 &  & +\left(b_{4}+\frac{1}{2}b_7\right)\epsilon^{AB}\epsilon^{ij}F_{AB}F_{ij}+b_{5}\epsilon^{AB}F_{Ai}F_{B}^{\ \ i}+b_{6}\epsilon^{ij}F_{Ai}F^{A}_{\ \ j},\label{a1}
\end{eqnarray}
where $G_{\mu\nu\rho}=\partial_\mu B_{\nu\rho}+\partial_\rho B_{\mu\nu}+\partial_\nu B_{\rho\mu}$ is the field strength for the Kalb-Ramond field. We only should remember that what would be the $B_{AB}$ and $B_{ij}$ components of $B_{\mu\nu}$ are set to zero from the onset, and then, $G_{ABi}=\partial_AB_{Bi}-\partial_BB_{Ai}$ and $G_{iBj}=\partial_iB_{Bj}-\partial_jB_{Bi}$. We could think the kinetic Lagrangian for the NG modes as a Lorentz-violating generalization of the Kalb-Ramond Lagrangian. In this perspective, we first imagine we break the $G_{\mu\nu\rho}G^{\mu\nu\rho}$ into $H$ representations with arbitrary parameters, and then we use the gauge invariance $B^\prime_{\mu\nu}=B_{\mu\nu}+\partial_\mu \xi_\nu-\partial_\nu \xi_\mu$ of $G_{\mu\nu\rho}$ to fix $B_{AB}=0$ and $B_{ij}=0$. In fact, the gauge freedom associated with the vector parameter $\xi_\mu$ enable us to fix three out of the six components of $B_{\mu\nu}$. Among the possible choices, we can, in particular, choose $\xi_A$ and $\xi_i$ to set $B_{AB}$ and $B_{ij}$ to zero. This still let us with the residual symmetry $B^\prime_{Ai}=B_{Ai}+\partial_A\xi_i-\partial_i\xi_A$, with $\partial_A\xi^A=-\partial_i\xi^i$ or, equivalently, with gauge symmetry (\ref{eq13}) of the Lagrangian (\ref{a1}). This gives us another way to motivate the invariance of (\ref{a1}) under the gauge symmetry (\ref{eq13}) as a condition for the linear stability of the model: we know the Kalb-Ramond gauge symmetry is needed to get rid of the longitudinal ghost modes of $B_{\mu\nu}$. Then, constructing an action for $B_{Ai}$, like in (\ref{eff action}), gives at the quadratic level (\ref{eq10}), which still allows the propagation of a dangerous residual longitudinal mode. The symmetry imposition (\ref{eq13}) exactly accounts for the inhibition of this mode.

Let us verify that, indeed, only one NG mode propagates in the model (\ref{a1}). A simple way to see this, is to work with the NG Lagrangian in the first order formalism. Defining two auxiliary fields, $\phi^A$ and $\phi^i$, we can write the NG Lagrangian as
\begin{equation}
\mathcal{L}_2^{NG} = 2a_{1}\left(\epsilon^{AB}\epsilon^{ij}G_{ABi}\phi_{j}-\frac{1}{2}\phi_i\phi^i\right)+2a_{2}\left(\epsilon^{AB}\epsilon^{ij}G_{iAj}\phi_{B}-\frac{1}{2}\phi_A\phi^A\right).\label{a3}
\end{equation}
The auxiliary fields do not have dynamics, and their equation of motion give simply
\begin{eqnarray}
\phi^i & = & \epsilon^{AB}\epsilon^{ji}G_{ABj},\label{a4}\\
\phi^A & = & \epsilon^{AB}\epsilon^{ij}G_{iBj}.\label{a5}
\end{eqnarray}
Plugging these solutions back into the Lagrangian (\ref{a3}), we regain the initial NG Lagrangian, proving the two models are equivalent.

Now, instead of substituting back (\ref{a4}) and (\ref{a5}) into (\ref{a3}), we derive the equation of motion for $B_{Ai}$. This gives
\begin{equation}
\epsilon^{BA}\epsilon^{ij}\left(a_{1}\partial_{B}\phi_{j}-a_{2}\epsilon^{BA}\phi_{B}\right)=0,\label{a6}
\end{equation}
whose solution can be conveniently written as
\begin{eqnarray}
\phi_{A}&=&\sqrt{\frac{a_1}{a_2}}\partial_A\chi, \label{a7}\\
\phi_{i}&=&\sqrt{\frac{a_2}{a_1}}\partial_i\chi.\label{a8}
\end{eqnarray}
Using equations (\ref{a6})-(\ref{a8}), we can put the Lagrangian (\ref{a3}) into the form
\begin{equation}
\mathcal{L}_2^{NG}=-a_{1}\partial_{A}\chi\partial^{A}\chi-a_{2}\partial_{i}\chi\partial^{i}\chi,\label{a9}
\end{equation}
which indeed describes just a massless scalar mode propagating according to the dispersion relation
\begin{equation}
p_Ap^A+\frac{a_2}{a_1}p_ip^i=0.\label{a10}
\end{equation}

We turn now to the investigation of the photon properties. In this case, we find more transparent to work directly with the equations of motion for the free photon field. From (\ref{a1}), these are given by
\begin{eqnarray}
2b_1\partial_AF^{AB}-b_3\partial_iF^{Bi}+\left(b_{4}+\frac{1}{2}b_7\right)\epsilon^{AB}\epsilon^{ik}\partial_AF_{ik}-b_{5}\epsilon^{BC}\partial_iF_{C}^{\ \ i}-b_{6}\epsilon^{ik}\partial_iF^{B}_{\ \ k}&=&0,\label{a15}\\
2b_2\partial_iF^{ij}+b_3\partial_AF^{Aj}+\left(b_{4}+\frac{1}{2}b_7\right)\epsilon^{AC}\epsilon^{ij}\partial_iF_{AC}+b_{5}\epsilon^{AC}\partial_AF_{C}^{\ \ j}+b_{6}\epsilon^{ji}\partial_AF^{A}_{\ \ i}&=&0. \label{a2}
\end{eqnarray}
It is convenient to restate these equations in terms of transverse and longitudinal components of the photon fields. Let us define:
\begin{eqnarray}
\epsilon^{AB}\partial_AA_B&=&\sigma_T,\label{a11}\\
\partial_BA^B&=&\sigma_L,\label{a12}\\
\epsilon^{ij}\partial_iA_j&=&\rho_T,\label{a13}\\
\partial_iA^i&=&\rho_L.\label{a14}
\end{eqnarray}

The quadratic Lagrangian (\ref{a1}) is invariant under the $U(1)$ gauge symmetry $A_A^\prime=A_A+\partial_A\alpha$ and $A_i^\prime=A_i+\partial_i\alpha$. Then, the transverse modes $\sigma_T$ and $\rho_T$ correspond to the true gauge invariant degrees of freedom of the photon. The gauge freedom allows, for example, to put $\rho_L$ or $\sigma_L$ to zero, and with the help of the equations of motion we can fix the other gauge degree of freedom in terms of the transverse modes.

Taking the rotational and divergence with respect to both derivative operators $\partial_A$ and $\partial_i$ of the equations (\ref{a15}) and (\ref{a2}), we get equivalent expressions in terms of the fields defined in (\ref{a11})-(\ref{a14})
\begin{eqnarray}
\left(b_1\partial_A\partial^A+\tilde{b}_3\partial_i\partial^i\right)\sigma_T+\tilde{b}_7\partial_B\partial^B\rho_T&=&0,\label{a16}\\
\left(\tilde{b}_4\partial_A\partial^A+b_2\partial_i\partial^i\right)\rho_T-\tilde{b}_7\partial_i\partial^i\sigma_T&=&0,\label{a17}\\
\left(\partial_A\partial^A\rho_L-\partial_i\partial^i\sigma_L\right)-\tilde{b}_5\partial_i\partial^i\sigma_T+\tilde{b}_6\partial_A\partial^A\rho_T&=&0\label{a18},
\end{eqnarray}
with
\begin{equation}
\tilde{b}_3=\frac{\left(b_3\right)^2-\left(b_5\right)^2}{2b_3},\ \ \tilde{b}_4=\frac{\left(b_3\right)^2+\left(b_6\right)^2}{2b_3},\ \ \tilde{b}_5=\frac{b_5}{b_3},\ \ \tilde{b}_6=\frac{b_6}{b_3},\ \ \tilde{b}_7=\frac{b_5b_6-b_3\left(2b_{4}+b_7\right)}{2b_3}.\label{a19}
\end{equation}

The two coupled equations (\ref{a16}) and (\ref{a17}) determine the dynamics of the transverse modes. We can easily see that the gauge freedom, together with the last equation, make the two longitudinal modes non-dynamical. If we choose $\alpha$ to make $\rho_L=0$, this condition, added to the requirement the function vanishes at spatial infinity, fixes completely the gauge freedom. With this choice, the last equation turns into a constraint equation for $\sigma_L$, which can be uniquely calculated in terms of the two dynamical transverse modes. We can also choose $\alpha$ to force $\sigma_L=0$. In this gauge, the last equation seems to give an extra propagating mode, but this choice only fix the gauge up to an extra function $\beta$ satisfying $\partial_B\partial^B\beta=0$, which can be used to constrain the homogeneous solution for $\rho_L$ to vanish, since this also satisfies $\partial_B\partial^B\rho_L=0$. Then, the solution for $\rho_L$ is given only in terms of the particular solution of the inhomogeneous equation (\ref{a18}).

To obtain the dispersion relations for the photon polarizations, we notice that the equations (\ref{a16}) and (\ref{a17}) only have nontrivial solutions provided
\begin{equation}
\text{det}\left(\begin{array}{cc}
b_1p_Ap^A+\tilde{b}_3p_ip^i & \tilde{b}_7p_Bp^B\\
-\tilde{b}_7p_ip^i & \tilde{b}_4p_Ap^A+b_2p_ip^i
\end{array}\right)=0,
\end{equation}
whose solutions are
\begin{equation}
\left(p_0^\pm\right)^2=\left(p_1\right)^2+u^\pm\left(\left(p_2\right)^2+\left(p_3\right)^2\right),\label{a20}
\end{equation}
where
\begin{equation}
u^\pm=\left(\frac{r}{2s}\pm\sqrt{\left(\frac{r}{2s}\right)^2-\frac{t}{s}}\right),\label{a21}
\end{equation}
and
\begin{equation}
r=\left(\tilde{b}_7\right)^2+\tilde{b}_3\tilde{b}_4+b_1b_2,\ \ s=b_1\tilde{b}_4,\ \ t=b_2\tilde{b}_3.\label{a22}
\end{equation}

Eq. (\ref{a20}) gives the desired dispersion relations for the photon polarizations.



\begin{thebibliography}{99}

\bibitem{Russell} V. A. Kostelecký and N. Russell, {\it Data Tables for Lorentz and CPT Violation},  Rev. Mod. Phys. $83$, $11$, (2011), arXiv:0801.0287.
\bibitem{lqg} J. Alfaro, H. A. Morales-Tecotl, and L. F. Urrutia, Phys. Rev. Lett. 84, 2318 (2000); Phys. Rev. D 65, 103509 (2002).
\bibitem{st} V. A. Kostelecký and S. Samuel, Phys. Rev. D 39, 683 (1989); Phys. Rev. Lett. 63, 224 (1989); Phys. Rev. Lett. 66, 1811
(1991); V. A. Kostelecký and R. Potting, Nucl. Phys. B359, 545 (1991); Phys. Lett. B 381, 89 (1996); Phys. Rev. D 63, 046007 (2001); V. A. Kostelecký, M. J. Perry, and R. Potting, Phys. Rev. Lett. 84, 4541 (2000).
\bibitem{wb} C.P. Burgess, J. Cline, E. Filotas, J. Matias, G.D. and Moore, JHEP 2002(03) 043 (2002), [hep-th/0201082].
\bibitem{ncg} See, for example, I. Mocioiu, M. Pospelov, and R. Roiban, Phys. Lett. B 489, 390 (2000); S.M. Carroll, J. A. Harvey,
V. A. Kostelecký, C. D. Lane, and T. Okamoto, Phys. Rev. Lett. 87, 141601 (2001); Z. Guralnik, R. Jackiw, S. Y. Pi, and A. P. Polychronakos, Phys. Lett. B 517, 450 (2001); C. E. Carlson, C. D. Carone, and R. F. Lebed, Phys. Lett. B 518, 201 (2001); A. Anisimov, T. Banks, M. Dine, and M. Graesser, Phys. Rev. D 65, 085032 (2002); A. Das, J. Gamboa, J. Lopez-Sarrion, and F. A. Schaposnik, Phys. Rev. D 72, 107702 (2005).
\bibitem{alan-ralf} V. A. Kostelecký and R. Lehnert, Phys. Rev. D 63, 065008 (2001).
\bibitem{carroll} S. M. Carroll, T. R. Dulaney, M. I. Gresham, and H. Tam, Phys. Rev. D 79, 065011 (2009).
\bibitem{Doplicher} S. Doplicher, K. Fredenhagen, and J. E. Roberts, {\it The Quantum structure of space-time at the Planck scale and quantum fields}, Commun. Math. Phys. 172, 187-224 (1995).
\bibitem{sme} D. Colladay and V. A. Kostelecký, Phys. Rev. D 55, 6760 (1997); Phys. Rev. D 58, 116002 (1998). V. A. Kostelecký, Phys. Rev. D 69, 105009 (2004).
\bibitem{alan-ncg} B. Altschul, Q. G. Bailey, and V. A. Kostelecký, Phys.Rev. D81 065028 (2010).
\bibitem{weinberg2} S. Weinberg, The Quantum Theory of Fields, Vol. II, Cambridge University Press, Cambridge, UK, (1996).
\bibitem{volkov} D. V. Volkov, Fiz. Elem. Chast. Atom. Yadra $3$ $(1973)$.
\bibitem{ogievetsky} V. I. Ogievetsky, "Nonlinear Realization of Internal and Space-time Symmetries", in Proc. of $10$th Winter School of Theoretical Physics in Karpacz, vol. $1$, Wroclaw $117 (1974)$.
\bibitem{penco} C. Armendariz-Picon, A. Diez-Tejedor, and R. Penco, JHEP 1010 079 (2010).
\bibitem{inverse-HE} E. Ivanov and V. I. Ogievetsky, Teor. Mat. Fiz. $25$ $164$ $(1975)$.
\bibitem{manohar} I. Low, A. V. Manohar, Phys. Rev. Lett. $88$ $(2002)$.
\bibitem{anisimov} A. Anisimov, T. Banks, M. Dine, and M. L. Graesser, Phys. Rev. D 65, 085032 (2002), [hep-ph/0106356].
\bibitem{Appelquist} T. Appelquist and J. Carazzone, Phys. Rev. D 11, 2856 (1975).
\bibitem{mukanov} E. Babichev, V. Mukhanov and A. Vikman, JHEP 0802, 101 (2008) [hep-th/0708.0561].
\bibitem{null} S. Dubovsky, T. Gregoire, A. Nicolis, and R. Rattazzi JHEP 0603 025 (2006).
\bibitem{arkani} A. Adams, N. Arkani-Hamed, S. Dubovsky, A. Nicolis, R. Rattazzi, JHEP 0610:014, (2006), [hep-th/0602178].
\bibitem{superluminal} Sergei Dubovsky and Sergey Sibiryakov, JHEP 0812 092 (2008).
\bibitem{wald} R. Wald, General relativity, The University of Chicago Press, (1984).


\end{thebibliography}
\end{document}